\documentclass[20pt,journal]{IEEEtran}
\usepackage{amsmath}
\usepackage{algorithm,algorithmic,amsbsy,amsmath,amssymb,epsfig,bbm,mathrsfs,fancyhdr,fancyvrb,subfigure,url}
\usepackage{amsfonts}
\usepackage{amssymb}
\usepackage{multirow}
\usepackage{bigstrut}
\usepackage{booktabs}
\usepackage{tabulary}
\usepackage{graphicx}
\usepackage{color,cite}
\usepackage{url}

\usepackage{pgfplots}
\usepackage{tikz}
\usepackage{graphicx,adjustbox}

\newenvironment{abbreviations}{\begin{list}{}{}}{\end{list}}

\usepackage{authblk}

\usepackage[justification=centering]{caption}

\usepackage{tikz}
\usetikzlibrary{mindmap}

\usepackage{breakurl}

\usetikzlibrary{patterns}

\newcommand{\ignore}[1]{}

\ifCLASSINFOpdf
\else
\fi

\makeatletter
\renewcommand\paragraph{\@startsection{paragraph}{4}{\z@}%
                                     {-3.25ex\@plus -1ex \@minus -.2ex}%
                                     {0.0001pt \@plus .2ex}%
                                     {\normalfont\normalsize\bfseries}}
\makeatother

\begin{document}
\title{Integration of Vehicular Clouds and Autonomous Driving: Survey and Future Perspectives \ignore{on Communications, Networking, Machine Learning, Computing, Security, and Autonomous Driving}}
\author{Yassine~Maalej,~\IEEEmembership{Student~Member,~IEEE,}
        and~Elyes~Balti,~\IEEEmembership{Student~Member,~IEEE} }
\maketitle
\begin{abstract}
For decades, researchers on Vehicular Ad-hoc Networks (VANETs) and autonomous vehicles presented various solutions for vehicular safety and autonomy, respectively. Yet, the developed work in these two areas has been mostly conducted in their own separate worlds, and barely affect one- another despite the obvious relationships. In the coming years, the Internet of Vehicles (IoV), encompassing sensing, communications, connectivity, processing, networking, and computation is expected to bridge many technologies to offer value-added information for the navigation of self-driving vehicles, to reduce vehicle on board computation, and to deliver desired functionalities. Potentials for bridging the gap between these two worlds and creating synergies of these two technologies have recently started to attract significant attention of many companies and government agencies. In this article, we first present a comprehensive survey and an overview of the emerging key challenges related to the two worlds of Vehicular Clouds (VCs) including communications, networking, traffic modelling, medium access, VC Computing (VCC), VC collation strategies, security issues, and autonomous driving (AD) including 3D environment learning approaches and AD enabling deep-learning, computer vision and Artificial Intelligence (AI) techniques.
We then discuss the recent related work and potential trends on merging these two worlds in order to enrich vehicle cognition of its surroundings, and enable safer and more informed and coordinated AD systems.  Compared to other survey papers, this work offers more detailed summaries of the most relevant VCs and ADs systems in the literature, along with some key challenges and insights on how different technologies fit together to deliver safety, autonomy and infotainment services.
\end{abstract}


\IEEEpeerreviewmaketitle

\textbf{ {\normalsize \\PART 1: Research Significance and Market Opportunity of Connected Vehicles and Self-Driving Cars\\} } \setcounter{section}{0}
The race among vehicle manufacturers and IT companies to develop, test and market a reliable fully autonomous vehicular technology is motivated by the big expected annual revenue stream of smart autonomous and connected vehicles. In this part, we introduce the major technological solutions in the IoV with their industrial market opportunities and research challenges and significance 

\section{Introduction and Research significance}{}
\IEEEPARstart{F}{}or more than three decades, researchers have increasingly infused more cyber capabilities to vehicles. Starting from basic communications to cell towers, passing through vehicle-to-vehicle (V2V) and vehicle-to-infrastructure (V2I) short range communications (a.k.a. VANETs), and reaching the recent era of VCs. On the other hand, many researchers have taken significant steps in vehicular automation with a long-term vision of reaching a full driving autonomy. Despite their obvious relevance to one another, the bridging and synergies between these two worlds have not been properly investigated. But this has recently started to attract massive interest from many companies and government agencies. For instance, the US National Science Foundation (NSF) and the US Department of Transportation (USDOT) have expressed tremendous need and importance to relate these two worlds together in several of their call for research proposals in 2017 \cite{IotSmartCarsChangingWorldForBetter}. They clearly emphasized the importance of integration and fusion of data from various input modes in order to create a deeper understanding of surroundings for safer and more coordinated AD applications. More precisely, enriched 3D scene reconstruction by different input technologies and deep learning techniques are of paramount importance to allow autonomous vehicle systems to perform effectively and safely. These directions are strongly supported by the few accidents and fatalities \cite{TeslaTragicLoss} and traffic light violations \cite{UberBlowsThroughRedLight} made by autonomous vehicle prototypes from top industries in the market (e.g., Tesla, Uber). These incidents that could have easily been mitigated if strong connectivity, computing and distributed coordination resources among vehicles and to the road infrastructure did exist.
Luckily, VANET applications for both safety and non-safety messaging have been approved by many government agencies (e.g., USDOT) and are “ready for deployment.” Forecasters estimate more than 250 million wireless-enabled vehicles by 2020 \cite{IotSmartCarsChangingWorldForBetter}. In addition, vehicle manufacturers, such as Cadillac, have already introduced commercial products (e.g., 2017 Cadillac CTS). Many modern vehicles will soon be equipped with sophisticated computing and storage capacities in their on-board unit (OBU). In addition, vehicles are getting more and more communication technology options, such as Dedicated Short Range Communications (DSRC), WiFi, 3G or Long-Term Evolution (LTE), which allow them to efficiently communicate and exchange information about their positions and intended steering commands. Furthermore, these wireless communication technologies can be used for awareness and safety through V2V beacons, as well as to enrich vehicular knowledge about their status and timings through V2I connections to static infrastructure entities such as traffic signals, street signs, etc.

In addition to the available computing capabilities, such as micro datacenters in modern RSUs as an example, modern vehicles are equipped with computing resources in their OBUs. All of these distributed resources together offer a great opportunity to form VCs. Advanced VC technologies efficiently combined with recent advances in AD  can revolutionize the latter. A VC does not only provide much more enriched and informed 3D understanding of the vehicle surroundings through communications among the vehicles and the infrastructure. But also migrates current AD decision-making paradigms from single isolated operations per vehicle to a  more cooperative and coordinated operation through VCC.
However, it is a difficult task to create a real-time 3D map that is enriched from the mapping of data derived from distinct sensing inputs as well as premapped data that usually provides static references on streets from the cloud. 
Reaching this stage requires not only merging and finding the synergies of VC and AD systems, but also getting separate data from the various technologies constituting each of them.  
For instance, the DSRC technology enables safety and non-safety applications for VANETs, which constitute a key component in VCs and ADs. Safety applications are time critical applications that are based on V2V communications that enrich the knowledge of AD systems, while non-safety applications rely mostly on V2I communications to enable the connectivity of vehicles to VCs.  
The Wireless Access for Vehicular Environment (WAVE) standard partitions the bandwidth into seven channels of 10 MHz each, one control channel (CCH) to serve safety applications, and six Service Channels (SCHs) to serve non-safety applications. DSRC specifies a channel switching scheme to allow vehicles to alternate between these two types of applications. Moreover, the channel switching scheme has a lot of limitations and needs to be improved and adjusted depending on the network density and the requirements of the applications.

Thus, there is a need for constructing robust single and multi hop communications for VCs in order to perform the desired coordinated role for multiple non-safety or commercial applications. This creates many challenges to optimize their coalition formation, manage the cloud resources efficiently and distributively, and make the technology more cost effective.

On the other hand, many deep learning and artificial intelligence systems and algorithms have been developed and tailored to enhance autonomous vehicles' 3D surrounding reconstruction and image flow recognition from both camera feeds and LIDAR point clouds. End-to-End recognition and driving decisions frameworks have also been recently studied. However, many challenges still exist in terms of the accuracy of the resulting learned environment and decisions taken by autonomous vehicles, especially in very low visibility (e.g., foggy weather, camera facing the sun) and out-of-sight scenarios (e.g., vehicles hidden by buildings/trucks at intersections), in which camera and LIDAR technologies can easily fail.

This article starts by providing a comprehensive survey, over-viewing the current research in the literature about both VC and AD technologies, and illustrating their on-going challenges and limitations. In addition, we shed some light on the recent work towards merging these two worlds to develop a synergistic framework for a more informed, advanced and coordinated AD systems. The outline of the contributions of this paper relative to the recent literature can be summarized as follows:
\begin{itemize}
    \item Present how different communications technologies and protocols fit together to form Vehicular networks and clouds. 
    \item Provide an overview of some of the VANET and VC challenges presented in the recent literature and provide a summary of the related research work. Moreover, we present the relation between these technologies and other emerging technologies, such as big data analytic and cloud computing.
    \item Unlike previous surveys, this survey provides not only the characteristics of the existing Media Access Control (MAC) protocols, but also how to optimize them to improve the performance of safety and non-safety applications in VCs.
    \item Due to the rapidly-changing topology and high node mobility in VANETs, we discuss different routing algorithms and their performance evaluation based on the transmission strategy, delay sensitivity and homogeneity or heterogeneity between networks. \item We classify various VCC architectures and expose their advantages and limitations. 
    \item We review various perceptions only, LIDAR only, and mixed approaches for objects recognition, tracking and scene flow estimation surrounding the self-driving vehicle.
    \item Review current research attempts ranging from pedestrian tracking, free space and road segmentation to more complex systems offering autonomous navigation based on the different up-to-date modalities of sensors.
    \item We present and compare recent Simultaneous localization and mapping (SLAM) based methods for vehicle localization based on surrounding on the move data recorded from different sensors.
    \item We present various open research and technical challenges of mapping and fusion techniques based on the sensing technology, alignment type (post or pre) and the data provenance from the VC or the vehicle sensors.
    \item We introduce recent works aiming to align the information obtained from camera feeds, LIDAR point clouds, and DSRC Basic Safety Messages (BSMs) in order to construct a more enriched and informed 3D understanding of autonomous vehicle surroundings.
    \item We shed light on prospective research directions to develop both more robust multimodal data merging and deep learning schemes for AD, and advanced computing and optimization frameworks for safer and more coordinated AD systems.
\end{itemize}
\section{Market Opportunity OF CONNECTED VEHICLES AND AD}
\begin{figure}[h] 
\begin{center} 
\includegraphics[width=9cm,height=12cm,keepaspectratio]{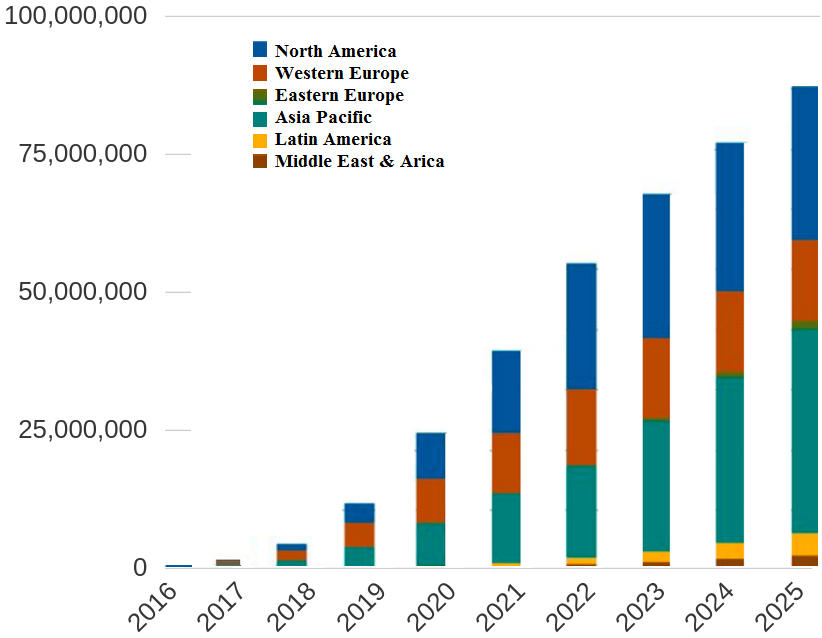}
\end{center}
\caption{Projected Numbers of Vehicles Equipped With V2X Connectivity Through 2025 \cite{ToyotaUmichV2XWorldCommunicationTest}.}
\label{DSRC_market}
\end{figure}

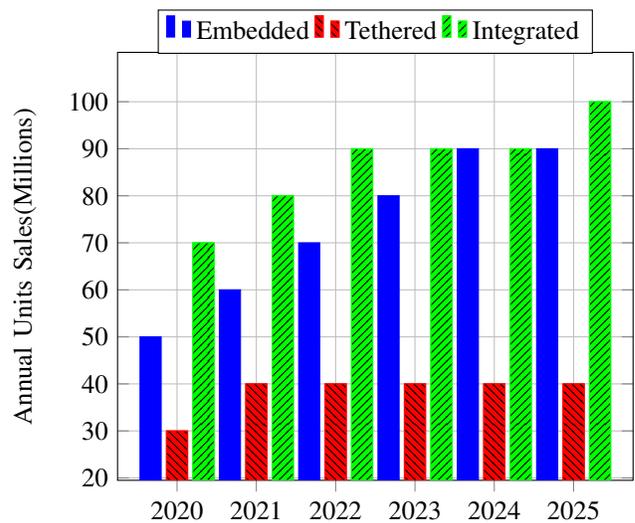
\begin{figure}[H]
\begin{tikzpicture}
\begin{axis}[
	x tick label style={
		/pgf/number format/1000 sep=},
	ylabel=Annual Units Sales(Millions),
	xlabel= \\$\quad$\\ $\quad$\\  ,
	symbolic x coords={2010, 2011, 2012, 2013, 2014, 2015, 2016, 2017, 2018,2019,2020,2021,2022,2023,2024,2025},
    xtick=data,
	enlargelimits=0.15,
	legend style={at={(0.5,1.1)},
		anchor=north,legend columns=-1},
	ybar,
	bar width=8pt,
	grid=major,
	symbolic y coords={0,10,20,30,40,50,60,70,80,90,100},
	ytick={0,10,...,100},
]
\addplot[ybar,color=blue,fill=blue] coordinates {
        (2020,50)
        (2021,60)
        (2022,70)
        (2023,80)
        (2024,90)
        (2025,90)
    };
\addplot[ybar,color=red,fill=red,postaction={pattern=north west lines}] coordinates {
        (2020,30)
        (2021,40)
        (2022,40)
        (2023,40)
        (2024,40)
        (2025,40)
        
    };
\addplot[ybar,color=green,fill=green,postaction={pattern=north east lines}] coordinates {
        (2020,70)
        (2021,80)
        (2022,90)
        (2023,90)
        (2024,90)
        (2025,100)
    };
\legend{Embedded, Tethered, Integrated}
\end{axis}
\end{tikzpicture}
\caption{Forecast Worldwide Sales of Vehicle Connectivity Solutions.}%
\label{WorldwideSalesForecastOfCarConnectivitySolutions}
\end{figure}

\begin{figure}[h] 
\begin{center} 
\includegraphics[width=9cm,height=12cm,keepaspectratio]{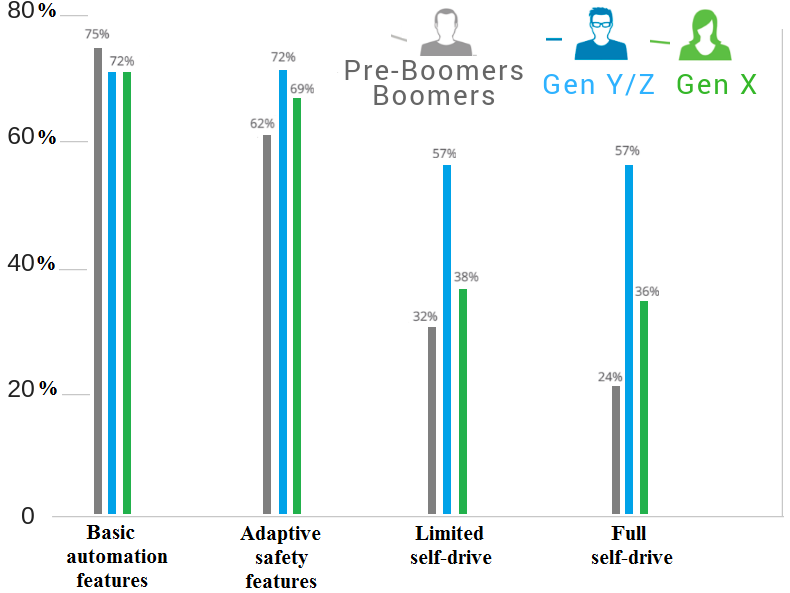}
\end{center}
\caption{US Consumer Interest of Various Levels of Vehicle Automation \cite{TrustAutonomousVehiclesTheRaceToAutonomousDriving}.}
\label{TrustAutonomousVehicles}
\end{figure}

{}
Connected and autonomous vehicles offer a promising market opportunity for on board communications equipment manufacturers, autonomous steering, 
Internet Service Providers (ISPs), LIDAR and RADAR scanners, AI vehicle computing platforms, etc. 
\begin{figure}[H] 
\begin{center} 
\includegraphics[width=7cm,height=7cm,keepaspectratio]{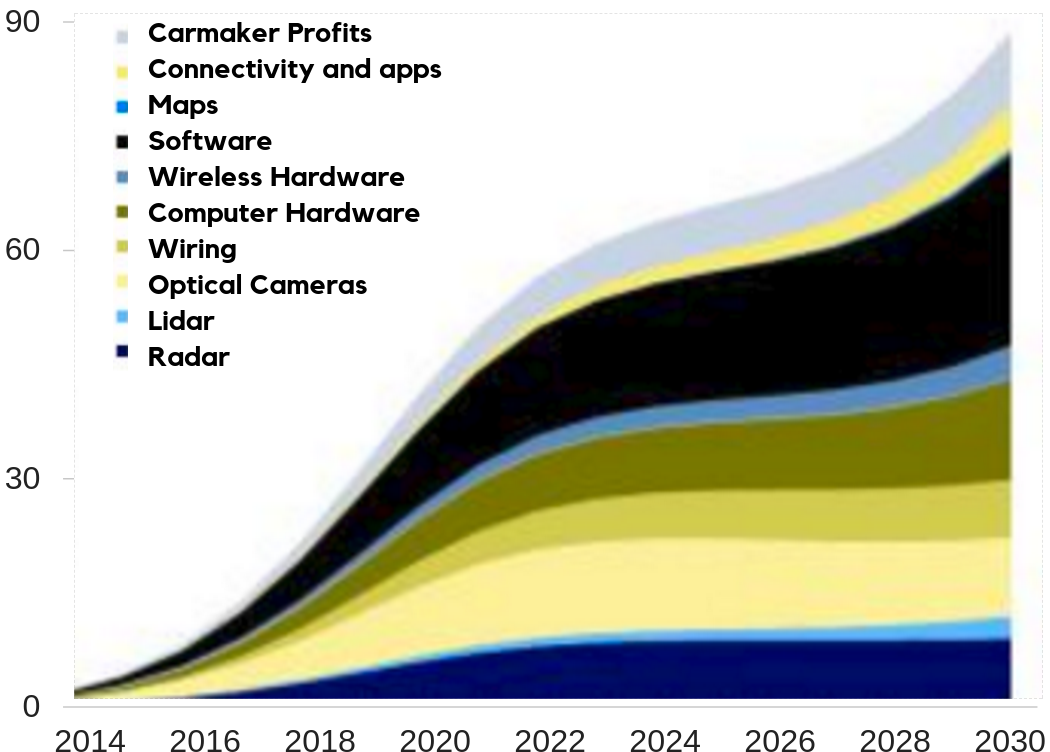}
\end{center}
\caption{Projected Market Share of Autonomous Vehicles Technology Through 2030 \cite{SelfDrivingCarsAn87BillionOpportunity}.}
\label{market_opportunity_autonomous_vehicles}
\end{figure} 
Appealed by the estimated \$87 billion market opportunity by 2030 \cite{SelfDrivingCarsAn87BillionOpportunity}, traditional vehicle manufacturers are thriving to put the pedal to the medal while they contest with technology companies to put driver assistive features of fully automated self-driving technology in vehicles. The software systems, computation hardware, and optical cameras are projected to form the most significant economic impact in the self-driving vehicles market, as presented in Fig. \ref{market_opportunity_autonomous_vehicles}.
The National Highway Traffic Safety Administration (NHTSA) has issued a notice of proposed rule-making (NPRM) to mandate vehicle manufacturers to mount DSRC radios for V2V communications in new vehicles sold in the USA market starting in 2020 \cite{NewCarsCouldBeRequiredToTalkToEachOtherAsSoonAs2020}. 
Consequently, this is expected to produce a rapid expansion of ubiquitous V2V and V2I connectivity, as the number of projections shown in Fig. \ref{DSRC_market} makes the leap to self-driving vehicles even smaller. Regardless of the on board communications, connectivity to the cloud has a substantial and profound impact on businesses and the society. It enables a variety of key vehicular applications such as software updates, real-time traffic with parking information, charging stations location and availability, and vehicular apps that evolve the deployment of autonomous vehicles.
Specifically by 2020, projections estimate that the connected vehicle market is expected to create a total revenue of \$141 Billion \cite{GlobalConnectedCarMarket}. In addition to the aforementioned driving related features, connected vehicles would gain from cloud services including advanced infotainment systems, commercial applications, and deep analysis and diagnostic systems. Based on different connectivity solutions, the annual global sales are expected to gain a tremendous growth especially for the embedded and integrated solutions, as shown in Fig. \ref{WorldwideSalesForecastOfCarConnectivitySolutions}.  The embedded connectivity solutions enable connectivity, intelligent computation and built in services and applications running inside the vehicle. While tethered solutions rely on the separation of the applications, remain inside the vehicle, and the connectivity is delivered externally (e.g., using a phone). The integrated solution (usually called bring-you-own-device) enables connecting a device that has the computation, applications and connectivity to the vehicle's dashboard.      
US consumer trust and interest in simple, partial, and fully automated driving systems have increased as detailed in Fig. \ref{TrustAutonomousVehicles} presented by the study in \cite{TrustAutonomousVehiclesTheRaceToAutonomousDriving}. Among all the levels of vehicle automation technologies, younger consumers in the US  have shown strong interest in systems with emphasis on basic automation, adaptive safety, limited autonomous driving and full autonomy. These survey findings presented more details about the trust and the interest of US customers between 2014 and 2016, which definitely point to a fast-pace growth demand towards the AD systems. 

However, this expected progression of spreading the connected VC related services and the AD systems is very related to the provision of high Quality of Service (QoS) connectivity for vehicles, and regulatory actions. 
\section{Remainder of the paper}
The overall structure of this article is organized into three major parts.
Part 2 provides a comprehensive survey of the key challenges encountered in VANETs and VCs. In Section \ref{SectionVehicularCommunicationTechnologies}, we introduce the various communications technologies used in vehicular networking. Heterogeneous VANETs are classified in Section \ref{SectionVehicularAdHocNetworks}. In Section \ref{SectionVanetMacProtocols}, we evaluate the numerous MAC protocols in VANETs. Different types and strategies for traffic modelings and networking of safety and non-safety applications are introduced in Section \ref{SectionTrafficModelingandNetworkingofSafetyandNon-SafetyApplications}. In addition, we present recent optimization schemes for MAC protocols and discuss the effect of the Synchronization Interval (SI) in the reliability of the vehicular applications. In Section \ref{SectionVehicularCloudComputing}, we compare the different ideas of VC formation based on various data center types and the heterogeneous communication technologies in ad-hoc cloud networks. We also show how differentiating the cloud performance can improve the QoS of the non-safety applications. Most common security threats in vehicular communications and VCs are described in Section IX.
Part 3 presents a thorough overview of the literature and the on-going open problems in AD and its enabling 3D scene reconstruction using camera and LIDAR technologies. In Section XII.A, we present mediated perception approaches for object recognition from cameras only. In Section XII.B, we present the prediction of driving decisions through direct perception approaches. While in Section XII.C we illustrate the behavior reflex approach usage to directly map sensed input to the driving action that has been previously taken. The various LIDAR based perception approaches for objects detection and recognition from a point cloud are developed in Section XII.D. The mixed techniques based on mediated and LIDAR perceptions are iterated in Section XII.E. Sections XIII.A,  XIII.B and XIII.C are dedicated to vehicle localization based on visual odometry, LIDAR and Combined Visual-LIDAR odometry, respectively.
Afterwards, Part 4 sheds light on the recent advances in merging these two worlds for advanced and coordinated AD systems. 
In Section XIV, recent techniques to post-align pre-learned scenes for camera feeds, LIDAR scans, VANET BSMs, and infrastructure messages are illustrated. Future research directions on pre-fusion of the data obtained from this multimodal input (i.e., camera, LIDAR and VANETs) are highlighted in Section XV. 
In addition, potential research tools that utilize the resulting enriched and informed environment recognition to achieve AD cooperation and coordination for AD are presented in Section XVI.
Finally, examples of integrating VCs and AD systems are presented in Section XVII.
In Part 5, we present the major open challenges facing the effective widespread utilization of VCs as well as the fully autonomous vehicles, as detailed in Section XVIII. Finally, we conclude the paper by summarizing the important take-away messages of this article in Section XIX.
\begin{center}
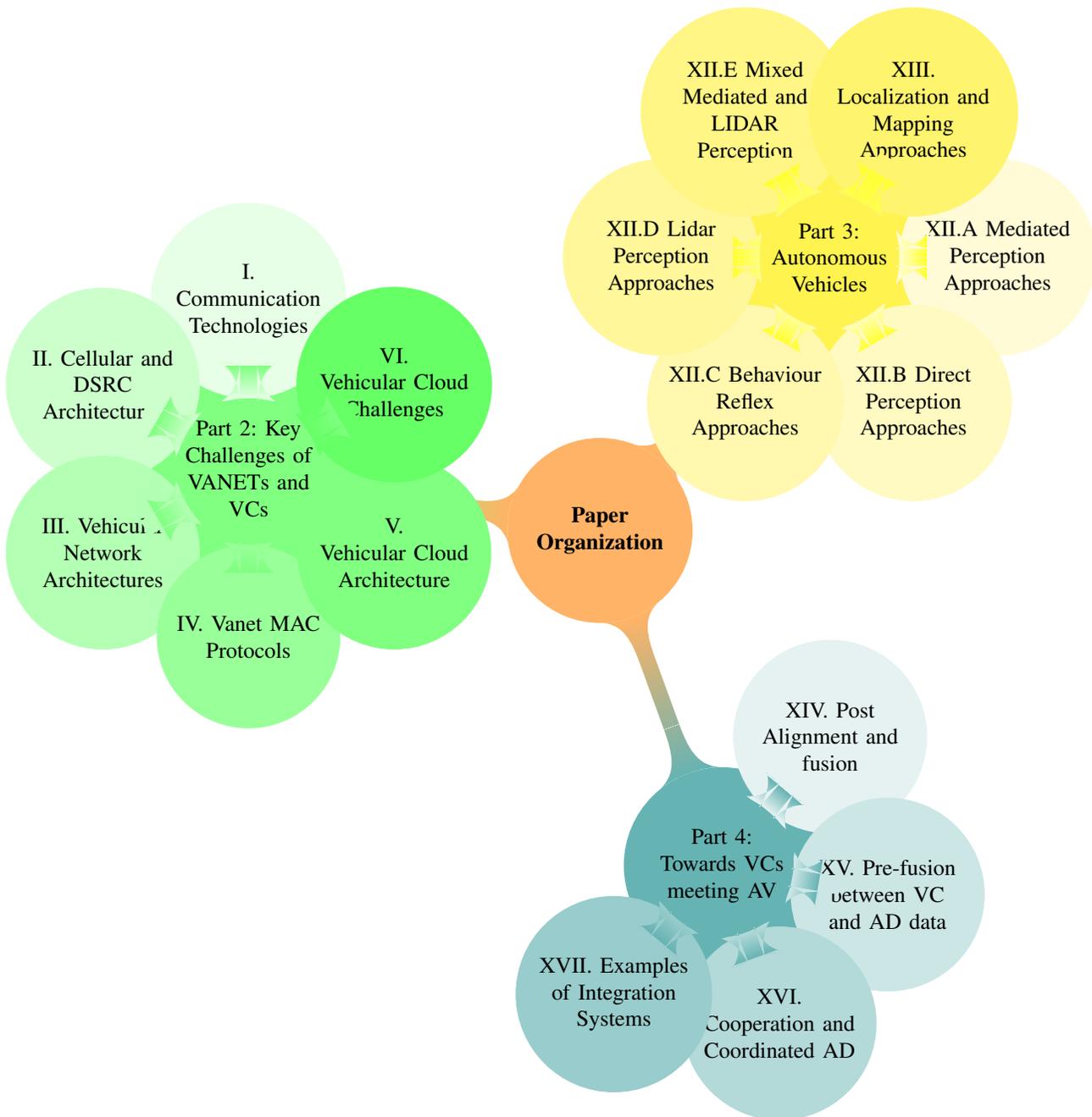
\begin{figure*}
\begin{tikzpicture}[scale=0.8, grow cyclic, text width=2.5cm, align=flush center, every node/.style=concept, 
concept color=orange!60, 
level 1/.style={level distance=7cm,sibling angle=120},
level 2/.style={level distance=3.3cm,sibling angle=60},
level 3/.style={level distance=4cm,sibling angle=0},
]
\node{\textbf{Paper Organization}} [clockwise from=50]  
    child [concept color=yellow!80] { node [concept] {Part 3: Autonomous Vehicles}  [clockwise from=60]
        child [concept color=yellow!10] { node {XII.A  Mediated Perception Approaches}}
        child [concept color=yellow!20] { node {XII.A  Mediated Perception Approaches}}
        child [concept color=yellow!30] { node {XII.B Direct Perception Approaches}}
        child [concept color=yellow!40] { node {XII.C Behaviour Reflex Approaches}}
        child [concept color=yellow!50] { node {XII.D Lidar Perception Approaches}}
        child [concept color=yellow!60] { node {XII.E Mixed Mediated and LIDAR Perception}}
        child [concept color=yellow!70] { node {XIII. Localization and Mapping Approaches}}
    }
   child [concept color=teal!60] { node {Part 4: \\Towards VCs meeting AV}
        child [concept color=teal!10] { node {XIV.  Post Alignment and fusion}}
        child [concept color=teal!20] { node {XV.  Pre-fusion \hspace{1cm}between VC\\ and AD data}}
        child [concept color=teal!30] { node {XVI. Cooperation and Coordinated AD}} 
        child [concept color=teal!40] { node {XVII. Examples of Integration Systems}}
    }
    child [concept color=green!50] { node {Part 2: Key Challenges of VANETs and VCs}  [counterclockwise from=90]
        child [concept color=green!10] { node {I.  \hspace{1cm} Communication Technologies}}
        child [concept color=green!20] { node {II. Cellular and DSRC Architectures}}
        child [concept color=green!30] { node {III.  Vehicular Network Architectures}}
        child [concept color=green!40] { node {IV. Vanet MAC Protocols}}
        child [concept color=green!50] { node {V. \hspace{1cm} Vehicular Cloud Architecture}}
        child [concept color=green!60] { node {VI. \hspace{1cm} Vehicular Cloud Challenges}}
    };
\end{tikzpicture}
\caption{Overall Organization of the Paper.}
\end{figure*}
\end{center}
\begin{center}
\textbf{ {\normalsize \\PART 2: KEY CHALLENGES IN VANETS AND VCS\\} } 
\end{center}
The major studied building blocks of vehicular networks can be classified into four major sections; namely: communications technologies, cellular and DSRC architectures, network architectures and VANET MAC protocols.
The vehicular networks are intended to manage the connectivity of vehicles from and to VCs and to support access independently of the clouds' architectures and challenges. This section focus on the overall challenges in VCs and provides a thorough summary of the most relevant architectures developed in the literature. \\
\section{Vehicular Communications Technologies}\label{SectionVehicularCommunicationTechnologies}
Modern vehicles are mostly equipped with various communications technologies such as WiFi, DSRC, 3G, and LTE. This is considered a valuable advancement, but ensuring that vehicles use the most appropriate communication technologies is a very challenging problem. This is the case for each of these technologies and the trade-off between their pros and cons. The maximum range of WiFi-direct can extend to 200 meters with a highest data rate of 11 Mbps. Whereas LTE is a paid mobile communications technology that has a wider connectivity coverage than DSRC and WiFi, but provides a lower data rate that ranges from 5 to 12 Mbps. Even though DSRC has a maximum data rate of up to 27 Mbps, its maximum range is 1000 meters. \\
\subsection{Dedicated Short-Range Communications (DSRC)}
\begin{figure}[h] 
\begin{center} 
\includegraphics[width=9cm,height=12cm,keepaspectratio]{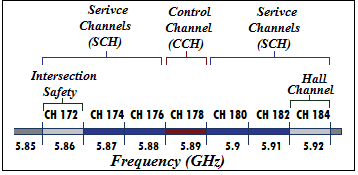}
\end{center}
\caption{DSRC Channel Arrangements \cite{YassineGlobecom}.}
\label{LabelWaveChannelArrangement}
\end{figure}

The DSRC \cite{DesignOf59GHzDSRCBasedVehicularSafetyCommunication} standard was coined by the Federal Communications Commission (FCC) \cite{FederalCommunicationsCommissionFCCReportAndOrderStd} to operate over a dedicated 75 MHz spectrum band around 5.9 GHz in the USA \cite{SafetyCommunicationsInTheUnitedStates} and partitions the bandwidth into seven channels. The DSRC has been adopted to enable two classes of applications (Safety and Non-Safety) for VANETs. One control channel CCH number 178 which are used to serve safety applications. Six SCHs (172,174,176,180,182,184) to serve non-safety applications as detailed in Fig. \ref{LabelWaveChannelArrangement}. 
\subsection{Wireless Access for Vehicular Environments (WAVE)}
\begin{figure}[h]
\begin{center} 
\includegraphics[width=9cm,height=8cm]{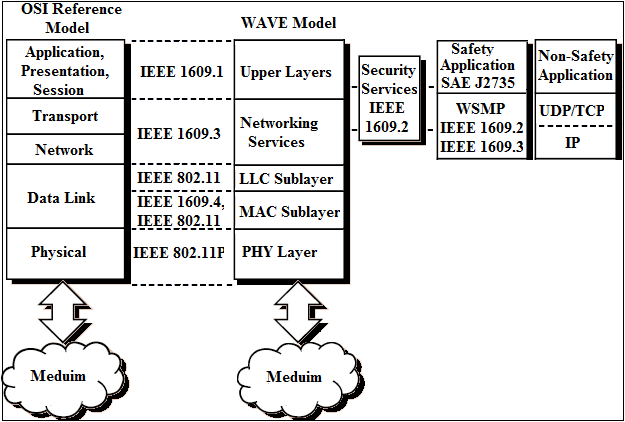}
\caption{OSI Model vs. WAVE Model Communications Stack \cite{YassineGlobecom}.}
\label{LabelOSIModelVsWAVEModelCommunicationStack}
\end{center}
\end{figure}

The CCH and the SCHs have four different access categories (ACs) and different priorities for each one of them to access the medium. In many urban areas, vehicles can use the free WiFi connectivity and might also use paid cellular base stations (e.g., 3G, LTE or 5G) connectivity to access the Internet depending on the application requirements, QoS, and cost. Vehicles having only DSRC connectivity may also use their neighboring vehicles as a gateway to the outside world. It is kind of tricky how to handle all these types of connections in a sophisticated and cost efficient manner.
In this section, we mainly focus on the DSRC emerging technology and WAVE standard of VANETs, then we describe different MAC protocols that have been proposed in many research attempts.
With respect to the OSI reference Model in Fig. \ref{LabelOSIModelVsWAVEModelCommunicationStack}, the WAVE model is divided into four categories: 
\begin{itemize}
    \item IEEE 1609.1 \cite{IEEEDraftGuideForWirelessAccessInVehicularEnvironmentsWAVEArchitecture}: standard used as a guide to manage the resources in VANET entities (e.g., RSUs, and OBUs).
    \item IEEE 1609.2: standard defined as the WAVE security services for applications and management of messages that are responsible for ensuring the anonymity, authenticity and confidentiality of the vehicular safety and non-safety packets. 
    \item IEEE 1609.3: standard developed as a guide for WAVE network configuration management and WAVE Short Message (WSM) transmission and reception.
    \item IEEE 1609.4 \cite{IEEE1609.4DSRCMultiChannelOperationsAndItsImplicationsOnVehicleSafetyCommunications}: standard responsible for WAVE Multi-Channel Operations that provides the DSRC frequency band with coordination and management of multiple connections by adopting the IEEE 802.11p standard for MAC and Physical Layer (PHY) specifications. 
\end{itemize}
\subsection{DSRC Channels Switching Assignment}
DSRC specifies a channel switching scheme that allows vehicles to alternate between the transmission of safety and non-safety packets. The DSRC standard recommends that, during a SI which is 100ms long, vehicles should visit the CCH to exchange their status messages with neighboring vehicles. The SI is composed of Control Channel Interval (CCHI) and Service Channel Interval (SCHI), both intervals are used to access the CCH and the SCHs for sending safety and non-safety application packets, respectively.
\begin{figure}[h]
\begin{center}
\includegraphics[width=9cm,height=12cm,keepaspectratio]{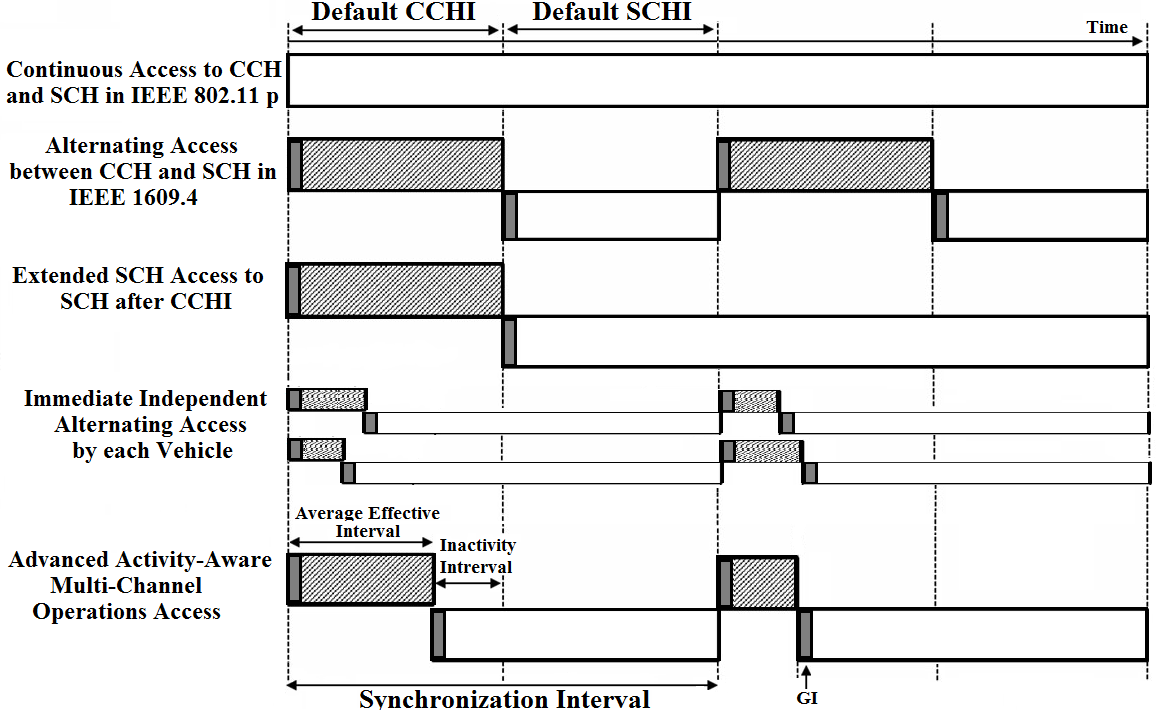}
\caption{WAVE Channel Access Mechanisms \cite{YassineGlobecom}.}
\end{center}
\end{figure}
Many research studies have been conducted to optimize the CCHI and the SCHI to guarantee a fair split of the SI between the safety and non-safety applications and to prioritize applications based on their QoS requirements.  A higher time share of the CCH from the SI leads to more reliable safety applications.
\begin{figure}[h]
\begin{center}
\includegraphics[width=9cm,height=12cm,keepaspectratio]{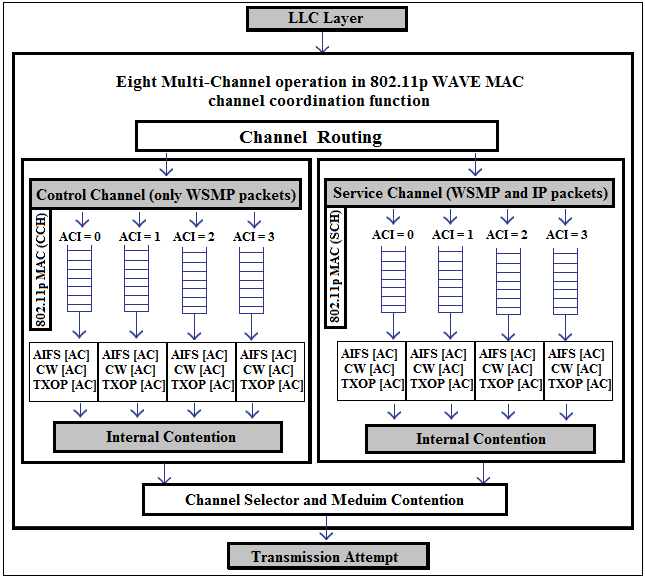}
\caption{Reference Model of MAC Architecture in IEEE 1069.4 for Multi-Channel Operations \cite{YassineGlobecom}.}
\end{center}
\end{figure}
In \cite{OptimizingtheControlChannelIntervaloftheDSRCforVehicularSafetyApplications}, the authors developed two algorithms to optimize the length of the CCHI so that non-safety applications have a fair share of the SI. In \cite{YassineGlobecom}, the authors developed an advanced awareness scheme of channel inactivity to dynamically achieve optimal interval switching based on the average effective interval utilization.
\section{Cellular Technology and DSRC-Cellular hybrid Architectures }\label{SectionCellularTechnology}
\begin{center}
    \begin{figure*} 
      \includegraphics[width=1.0\textwidth]{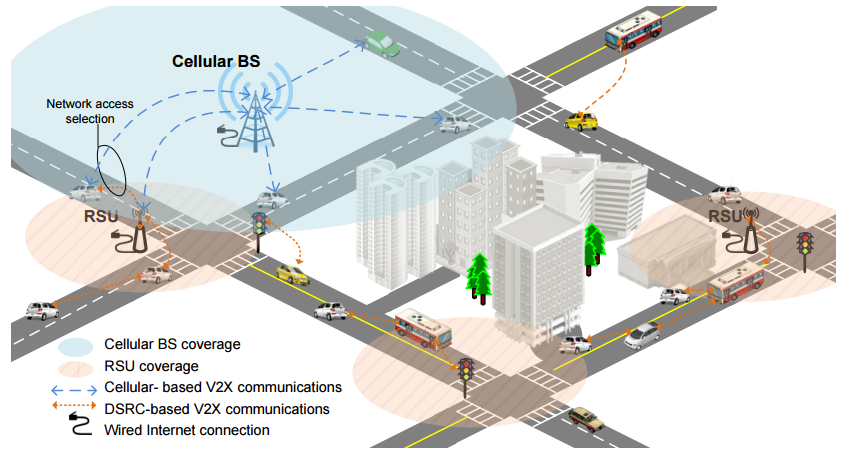}
      \caption{DSRC-Cellular Hybrid Scenario for V2X Communications \cite{InterworkingOfDSRCAndCellularNetworkTechnologiesForV2XCommunicationsASurvey}.}
      \label{DSRCCellularHybridScenarioForV2XCommunications}
    \end{figure*}
\end{center}

Due to the aforementioned limitations of DSRC, many research attempts have raised interest to investigate the support of cellular technologies (e.g., 3G, LTE, 5G) for vehicle to any (V2X) communications and usage of both DSRC and cellular communications, as shown in Fig. \ref{DSRCCellularHybridScenarioForV2XCommunications}. Cellular technology is gaining momentum in the research community over DSRC because of the following enablers: 
\begin{itemize}
    \item The capacity of dense networks to support high bandwidth demand of vehicles and cellular users at the same time. In such scenario, the performance evaluation of the downlink with unicast and broadcast vehicular messages is studied in \cite{InterVehicleCommunicationAtIntersectionsAnEvaluationOfAdHocAndCellularCommunication} for LTE based systems.
    \item Reduced frequency of horizontal handoff of vehicles between BSs is guaranteed through the large cellular coverage.
    \item Unlike DSRC poor Non-Line-Of-Sight receptions, cellular LTE systems provide potentially better coverage as demonstrated in \cite{AComparisonOfUMTSAndLTEForVehicularSafetyCommunicationAtIntersections}. 
    \item Cellular is widespread and mature technology that accelerates the deployment of V2X communications and does not require new infrastructure facilities like DSRC.
\end{itemize}
On the contrary, current research studies that investigate sending vehicular safety messages using LTE \cite{SafetyOnTheRoadsLTEAlternativesForSendingITSMessages}, and cross-traffic assistance of vehicular safety messages using UMTS and LTE \cite{AComparisonOfUMTSAndLTEForVehicularSafetyCommunicationAtIntersections}, do not account for traditional cellular network traffic. Even though the downlink and uplink transmission rates between the BS, as the access network connecting to the Internet, and vehicles in LTE (up to 300 and 75 Mb/s, respectively) are higher than in DSRC, LTE cannot provide the same awareness update rate and latency for Cooperative Awareness Messages (CAM) like "offered" using DSRC \cite{AComparisonOfUMTSAndLTEForVehicularSafetyCommunicationAtIntersections}. To put it differently, the centralized nature of cellular networks imposes that messages should be sent from vehicles to BSs that unicast or broadcast messages to other vehicles. In other words, instead of directly sending safety messages using DSRC, in cellular networks safety messages have to pass by the BS. Therefore, the use of the cellular technology results in lower awareness update rate and higher latency for disseminating delay critical safety-related applications. Additionally, several challenges limit transmission of the messages received by the BS:
\begin{itemize}
    \item Unicasts the received safety messages to the relevant vehicles subject to the the initial sender's safety. The Work in \cite{LTEForVehicularNetworkingASurvey} shows that in case of unicast of messages from the BS, the uplink has lower traffic load than the downlink and it becomes a bottleneck. One affordable solution that is available in 3GPP and higher standards is the use of multimedia broadcast and multicast services (MBMS), and the evolved MBMS (eMBMS) for safety message dissemination over cellular communications.
    \item Broadcasts the received safety messages to all the vehicles in its cell coverage area. Ultimately every vehicle receives all the safety messages sent by the BS, regardless of whether or not it belongs to the same zone of relevance to the sending vehicle. An alternative solution that might have a high latency is to create multicast groups \cite{MultimediaBroadcastMulticastServiceMBMSArchitectureAndFunctionalDescriptionRelease13} with eMBMS that are concerned with the safety messages that the BS transmits. 
\end{itemize}
Detailed comparisons between the DSRC-cellular and pure cellular architectures are summarized in Table. \ref{CellularDSRCHybridClassification} 
\begin{table*}[t]
\centering
\caption{PURE CELLULAR AND HYBRID DSRC-CELLULAR ARCHITECTURES FOR V2X COMMUNICATIONS}
\label{CellularDSRCHybridClassification}
\begin{tabular}{lllllll}
\hline
\multicolumn{1}{|c|}{Architecture}                                                                                 & \multicolumn{1}{l|}{\begin{tabular}[c]{@{}l@{}}Channel\\ Seperation\end{tabular}} & \multicolumn{1}{l|}{\begin{tabular}[c]{@{}l@{}}Channel\\ Assignment\end{tabular}}                    & \multicolumn{1}{c|}{Reference}                                                                   & \multicolumn{1}{c|}{Application}                                                        & \multicolumn{1}{l|}{\begin{tabular}[c]{@{}l@{}}Vehicular\\ Communication\end{tabular}} & \multicolumn{1}{l|}{\begin{tabular}[c]{@{}l@{}}Bandwidth\\ Aggregation\end{tabular}}           \\ \hline
\multicolumn{1}{|c|}{\multirow{2}{*}{\begin{tabular}[c]{@{}c@{}}Pure Cellular\\ Architecture\end{tabular}}}        & \multicolumn{1}{c|}{\multirow{2}{*}{Hierarchical}}                                & \multicolumn{1}{c|}{Fixed}                                                                           & \multicolumn{1}{c|}{\cite{FixedChannelAssignmentAndLayerSelection}}                                                                    & \multicolumn{1}{c|}{Any application}                                                               & \multicolumn{1}{c|}{Cellular}                                                              & \multicolumn{1}{c|}{No Aggregation}                                                                      \\ \cline{3-7} 
\multicolumn{1}{|c|}{}                                                                                             & \multicolumn{1}{c|}{}                                                             & \multicolumn{1}{c|}{Dynamic}                                                                       & \multicolumn{1}{c|}{\cite{DynamicChannelAssignmentStrategiesInCellularMobile}}                                                                 & \multicolumn{1}{c|}{Any application}                                                               & \multicolumn{1}{c|}{Cellular}                                                              & \multicolumn{1}{c|}{No aggregation}                                                                      \\ \hline
\multicolumn{1}{|c|}{\multirow{5}{*}{\begin{tabular}[c]{@{}c@{}}DSRC-Cellular\\ Hybrid Architecture\end{tabular}}} & \multicolumn{1}{c|}{\multirow{3}{*}{Flat}}                                        & \multicolumn{1}{c|}{\multirow{2}{*}{\begin{tabular}[c]{@{}c@{}}Performance \\ Centric\end{tabular}}} & \multicolumn{1}{l|}{\cite{HeterogeneousVehicularWirelessNetworkingATheoreticalPerspective}}    & \multicolumn{1}{c|}{\begin{tabular}[c]{@{}c@{}}Internet \\ Access\end{tabular}}         & \multicolumn{1}{c|}{\begin{tabular}[c]{@{}c@{}}V2I\\ No V2V\end{tabular}}              & \multicolumn{1}{c|}{\begin{tabular}[c]{@{}c@{}}Yes \\ Generic/DSRC\\ Aggregation\end{tabular}} \\ \cline{4-7} 
\multicolumn{1}{|c|}{}                                                                                             & \multicolumn{1}{c|}{}                                                             & \multicolumn{1}{c|}{}                                                                                & \multicolumn{1}{l|}{ \cite{ProperHandoverBetweenVANETAndCellularNetworkImprovesInternetAccess}} & \multicolumn{1}{c|}{\begin{tabular}[c]{@{}c@{}}Internet \\ Access\end{tabular}}         & \multicolumn{1}{c|}{\begin{tabular}[c]{@{}c@{}}V2I\\ No V2V\end{tabular}}              & \multicolumn{1}{c|}{\begin{tabular}[c]{@{}c@{}}No\\ LTE/DSRC\\ Aggregation\end{tabular}}       \\ \cline{3-7} 
\multicolumn{1}{|c|}{}                                                                                             & \multicolumn{1}{c|}{}                                                             & \multicolumn{1}{c|}{Data-centric}                                                                    & \multicolumn{1}{l|}{\cite{ImprovementOfVehicularCommunicationsByUsing3GCapabilitiesToDisseminateControlInformation}, \cite{TowardsSoftwareDefinedVANETArchitectureAndServices}}                                                                            & \multicolumn{1}{c|}{Routing}                                                            & \multicolumn{1}{c|}{\begin{tabular}[c]{@{}c@{}}V2V\\ No V2I\end{tabular}}              & \multicolumn{1}{c|}{\begin{tabular}[c]{@{}c@{}}No\\ UMTS/DSRC\\ Aggregation\end{tabular}}      \\ \cline{2-7} 
\multicolumn{1}{|c|}{}                                                                                             & \multicolumn{1}{c|}{\multirow{2}{*}{Hierarchical}}                                & \multicolumn{1}{l|}{Dynamic}                                                                         & \multicolumn{1}{l|}{\cite{QoEEnhancementOfSVCVideoStreamingOverVehicularNetworksUsingCooperativeLTE802.11pCommunications}, \cite{DynamicClusteringBasedAdaptiveMobileGatewayManagementInIntegratedVANETx20143GHeterogeneousWirelessNetworks}}                                                                            & \multicolumn{1}{c|}{Video Sharing}                                                      & \multicolumn{1}{c|}{V2V and V2I}                                                       & \multicolumn{1}{c|}{\begin{tabular}[c]{@{}c@{}}Yes \\ LTE/DSRC \\ Aggregation\end{tabular}}    \\ \cline{3-7} 
\multicolumn{1}{|c|}{}                                                                                             & \multicolumn{1}{c|}{}                                                             & \multicolumn{1}{c|}{Fixed}                                                                           & \multicolumn{1}{l|}{\cite{CloudAssistedSafetyMessageDisseminationInVANETCellularHeterogeneousWirelessNetwork}}                                                                            & \multicolumn{1}{c|}{\begin{tabular}[c]{@{}c@{}}Safety Message\\ Broadcast\end{tabular}} & \multicolumn{1}{c|}{V2V and V2I}                                                       & \multicolumn{1}{c|}{\begin{tabular}[c]{@{}c@{}}No\\ LTE/DSRC \\ Aggregation\end{tabular}}      \\ \hline
            
\end{tabular}
\end{table*}
\section{Vehicular Network Architectures}\label{SectionVehicularAdHocNetworks}
In this section, we survey the most recent research progress on vehicular network architectures, as detailed in \cite{RoutinginvehicularadhocnetworksAsurvey}. This can be classified into three major categories based on the parts intervening in the network:
\begin{itemize}
    \item Stationary cellular and/or RSU based networks used as access points to provide services to vehicles, such architecture is detailed in \cite{EngineeringLinkUtilizationInCellularOffloadingOrientedVANETs}. 
    \item Pure VANET formed by parked \cite{TowardsFaultTolerantJobAssignmentInVehicularCloudComputing} or moving vehicles \cite{ASurveyOfMobileCloudComputingArchitectureApplicationsAndApproaches}.
    \item Hybrid network architecture that combines cellular and RSU with mobile vehicles as ad-hoc networks, as described in \cite{AStudyOnTheFeasibilityOfMobileGatewaysForVehicularAdHocNetworks}. 
\end{itemize}
Vehicular networks based on fixed infrastructure are eventually unfeasible because of the high cost associated with infrastructure deployment. Most of the research attempts focus on VANETs and hybrid networks. VANETs are created by applying the principles of mobile ad-hoc networks (MANETs) for spontaneously creating a wireless network of moving vehicles to increase safety through DSRC communications or collaboration to better utilize the communications capabilities of vehicles in the same area. VANETs support a wide range of applications that vary from simple one hop \cite{ReliabilityAnalysisOfOneHopSafetyCriticalBroadcastServicesInVANETs} information dissemination of CAMs \cite{AnApproachForSelectiveBeaconForwardingToImproveCooperativeAwareness} to multi-hop \cite{RoutinginvehicularadhocnetworksAsurvey} dissemination of messages over vast distances \cite{Anewbroadcastprotocolforvehicularadhocnetworkssafetyapplications}. VANETs can be divided into three types depending on their building blocks; namely, connected vehicles that rely only on V2V communications, coalition of RSUs and OBUs with V2V and V2I communication, and interconnected infrastructure equipment (e.g., Traffic light, pavement markers, cameras, etc.). In contrast to MANETs, modeling the moving nodes in VANETs is less complex especially that most of vehicles are restricted in their range of motion by paved highways/roads with traffic lights. The key issue of network selection is when vehicles are connected to many types of infrastructure nodes like RSUs with DSRC communications access and/or cellular nodes with 3G/LTE access and/or WiFi. Thus, it renders the decision making regarding the best type of connectivity very critical given vehicular application requirements, cost, availability expectations, etc. These infrastructure nodes have to be utilized efficiently in the best manner that reduces handover latency from one network to another and to avoid bottleneck network congestion. In \cite{AnIPv6ArchitectureForCloudtoVehicleSmartMobilityServicesOverHeterogeneousVehicularNetworks}, authors studied the use of IPv6 based on its efficient Internet traffic flow management over heterogeneous technologies. Their model is based on using the RSUs to act as IPv6 routers and to handle the convergence of DSRC and cellular technologies while vehicles get in or out of their coverage. 
\section{VANET MAC PROTOCOLS} \label{SectionVanetMacProtocols}
A major challenge in VANETs is the frequent changes in the network topology. VANET MAC protocols have to reduce the medium access delay for safety applications and to handle all the vehicles joining or leaving the network. In this paper, we classify several MAC protocols that have been proposed for VANETs into five major categories; namely, contention-based protocols, contention-free protocols, hybrid protocols, dedicated short-range communications-based protocols, and directional antenna-based protocols. In VANETs, Time Division Multiple Access (TDMA), Frequency Division Multiple Access (FDMA), or Code Division Multiple Access (CDMA) are very difficult to implement because of the need to dynamically allocate slots, channels, or codes without centralized control.
\begin{table*}[t]
\begin{center}
\caption{WAVE MAC PROTOCOLS CLASSIFICATION}
\begin{tabular}{|l|l|l|l|}
\hline
Channel Access & Channel Separation & Channel Assignment & Channel Coordination  \\ 
\hline
\multirow{5}{*}{Hybrids} & 
  \multirow{2}{*}{Contention-free methods} 
  & Distributed & \verb|    ***| \\ \cline{3-4}
  & & \multirow{2}{*}{Centralized RSU} & \verb|ACFM| \cite{AnadaptivecollisionfreeMACprotocolbasedonTDMAforintervehicularcommunication10} \\ \cline{4-4}   & & &  \verb|CBRC| \cite{RSUcentricchannelallocationinvehicularad-hocnetworks97} \\ \cline{2-4}
  
  & \multirow{7}{*}{\parbox{2cm}{Contention-(free and based) methods}}
  & \multirow{3}{*}{Distributed} & \verb|CS-TDMA |\cite{AhybridefficientandreliableMACforvehicularadhocnetworks77} \\ \cline{4-4}   & & & \verb|SOFT-MAC| \cite{SOFTMACSpaceorthogonalfrequencytimemediumaccesscontrolforVANET54} \\ \cline{4-4} & & &  \verb|DMMAC| \\ \cline{4-4} & & &  \verb|HER-MAC| \\ \cline{3-4} 
  & & \multirow{2}{*}{Centralized Cluster-head} 
  & \verb|CBMMAC| \\ \cline{4-4}  & & &   \verb|CBMCS| \\ \cline{3-4}
  & & \multirow{1}{*}{Centralized RSU} 
  & \verb|RMAC| \\ \cline{2-4}
\hline \multirow{3}{*}{Contention-Based} 

 & \multirow{2}{*}{IEEE 802.11p \cite{IEEEStandardForInformationTechnologyTelecommunications}} 
 &  \multirow{2}{*}{Distributed} & \verb|CSMA with DEA| \cite{IEEE80211pPerformanceEvaluationAndProtocolEnhancement} \\ \cline{4-4}  
 & & &  \verb|DCAS| \cite{AnIEEE80211pBasedDistributedChannelAssignmentSchemeConsideringEmergencyMessageDissemination} \\ \cline{3-4}
 & & Centralized RSU & \verb|p-persistant CSMA with CEA| \cite{IEEE80211pPerformanceEvaluationAndProtocolEnhancement} \\ \cline{2-4}
\hline
\multirow{5}{*}{Contention-free} 
  & \multirow{7}{*}{\parbox{2cm}{TDMA}}
  
  & \multirow{7}{*}{Distributed} 
    & \verb|VeMAC| \cite{VeMACATDMAbasedMACprotocolforreliablebroadcastinVANETs} \\ \cline{4-4}   
    & & & \verb|VeSOMAC| \cite{AselfreorganizingMACprotocolforintervehicledatatransferapplicationsinvehicularadhocnetworks}  \\ \cline{4-4}  
    & & & \verb|STDMA| \cite{STDMAforvehicletovehiclecommunicationinahighwayscenario}  \\ \cline{4-4} 
    & & &  \verb|CFR MAC| \\ \cline{4-4}
    & & &  \verb|ASTA| \\ \cline{3-4} 

& & \multirow{2}{*}{Centralized Cluster-head}
    &      \verb| CAH-MAC|\cite{CAHMACCooperativeadhocMACforvehicularnetworks} \\ \cline{4-4} 
    & & &  \verb|CBMAC|\cite{ClusterBasedMediumAccessSchemeForVANETs} \\ \cline{4-4}
    & & &  \verb|ADHOC MAC|\cite{ADHOCMACNewMACarchitectureForAdHocNetworksProvidingEfficientAndReliablePointToPointAndBroadcastServices}\\\cline{4-4}
    & & &  \verb|A-ADHOC |\cite{AADHOCAnAdaptiveRealTimeDistributedMACProtocolForVehicularAdHocNetworks}\\\cline{4-4}
    & & &  \verb|TC-MAC|\cite{TDMAClusterBasedMACForVANETsTCMAC}\\\cline{4-4}
    & & &  \verb|CBT|\cite{AclusterbasedTDMASystemForIntervehicleCommunications}\\\cline{3-4}

  & & \multirow{1}{*}{Centralized RSU}
  &  \verb|UTSP| \cite{AnovelCentralizedTDMAbasedSchedulingProtocolForVehicularNetworks} \\ \cline{2-4}
& \multirow{1}{*}{\parbox{2cm}{FDMA}}
& \multirow{1}{*}{Distributed} & \verb|dFDMA|\cite{ANewApproachToExploitMultipleFrequenciesInDSRC} \\  \cline{3-4}
\hline
\end{tabular}
\end{center}
\end{table*}
We highlight the MAC protocols using TDMA by implementing time-slotted access channels while taking into consideration the major requirement of clock synchronization between vehicles. TDMA allocates the disposed bandwidth to all the vehicles by creating different time frames and dividing each frame into several time slots so that vehicles send their data during one or more time slots. Vehicles can access every frame and they receive during the time slots allocates to neighboring vehicles. The rapidly changing topology in VANETs is induced by the fast moving vehicles, which imposes that an efficient MAC protocol must be implemented to adapt to these changes in the network either through the use of specialized RSUs, or using cluster head vehicles, or in a distributed manner.
In contrast to TDMA time slotted accesses, MAC protocols developed as part of the IEEE 802.11p standard offer a significant increase in the network throughput while reducing the transmission delay by using CCH and SCHs with reservation-based access and dynamic assignment of channels to resolve the hidden terminal problem \cite{MultiChannelMACForAdHocNetworksHandlingMultiChannelHiddenTerminalsUsingASingleTransceiver}.  Ko \textit{et al.} \cite{MediumAccessControlProtocolsUsingDirectionalAntennasinAdHocNetworks} developed a directional antenna-based MAC protocol that can be applied to vehicular environments so that it allows a vehicle to block the communication to avoid the hidden node problem. 
\section{Network Selection Schemes and Routing Protocols in VANETs}\label{SectionRoutingProtoclsInVanets}
The dynamic nature of nodes in VANETs offers additional challenges of finding and maintaining routes for vehicular application packets. In addition, the high mobility induces difficulties while routing in homogeneous or heterogeneous networks. The routing challenges in VANETs include: 
\begin{itemize}
  \item Routing in Homogeneous Networks: most of the traditional VANET routing protocols have one common assumption that all packets are transmitted via a short distance wireless technology (e.g., WAVE standards). As a result, those routing algorithms' performance is often analyzed and compared relative to a horizontal handoff. The work in \cite{VehicularTelematicsOverHeterogeneousWirelessNetworksASurvey} presents the scenario of a data transmission session transferred from one Point of Attachment (POA) to another with the same access technology and in the same network.
In \cite{ARobustHandoverUnderAnalysisOfUnexpectedVehicleBehaviorsInVehicularAdHocNetwork} and \cite{MobilityAndHandoffManagementInVehicularNetworksASurvey}, the authors studied a use case of horizontal handover to transfer an Internet video streaming session of a vehicle while moving from one coverage area of an RSU to another. 
\item Routing in Heterogeneous Networks: The main reason why IoV networks are more complicated than VANETs is that they often include different radio access technologies, thereby leading to a heterogeneous network. Vertical handoff is a classical problem in heterogeneous networks, including the IoV, with different access technologies that have different characteristics (e.g., bandwidth, frequency band, modulation and coding scheme). It offers the possibility to switch from one access technology to another based on performance, availability or cost reasons, while maintaining active connections. The authors in \cite{VerticalMobilityManagementArchitecturesInWirelessNetworksAComprehensiveSurveyAndFutureDirections} present the scenario of a data transmission session transferred from one POA to another with a different access technology and in a different network.  
In \cite{VerticalHandoffDecisionAlgorithmsForProvidingOptimizedPerformanceInHeterogeneousWirelessNetworks}, Lee \textit{et al.} proposed a vertical handoff (VHO) algorithm for more seamless integration of 3/4G wireless data networks, VANETs and IEEE 802.11 WLANs.  
\end{itemize}
\subsection{Network Selection Schemes}
One way to increase the vehicular networks performance in terms of throughput and delay is to minimize the cost related to handovers. Network selection schemes are effective processes to make handover decisions depending on handover triggers. They are considered as measures to derive the handover decisions and can be classified into user-centric, network-centric or hybrid, as illustrated in details in Table \ref{TableHandoverTriggers}. In cases where there is a necessity for network improvement or handover needs to be applied, user and network related information are provided to the network selection scheme in order to make the network selection's decision be more flexible. 



\begin{table*}[t]
\centering
\caption{HANDOVER TRIGGERS AND SELECTION SCHEMES}
\label{TableHandoverTriggers}
\begin{tabular}{|c|c|c|l|}
\hline
Handover Trigger & Type & Metrics & \multicolumn{1}{c|}{References} \\ \hline
\multirow{2}{*}{User-centric} & \begin{tabular}[c]{@{}c@{}}QoS Provisioning\end{tabular} & \begin{tabular}[c]{@{}c@{}}End to end delay and packet loss \\ eg. Quality of Video  Streaming\end{tabular} & \begin{tabular}[c]{@{}l@{}}\cite{UserCentricMobilityManagementArchitectureForVehicularNetworks} ,  \cite{QoEEnhancementOfSVCVideoStreamingOverVehicularNetworksUsingCooperativeLTE802.11pCommunications}\end{tabular} \\ \cline{2-4} 
   & Cost & \begin{tabular}[c]{@{}c@{}}Cellular services are provided\\ with subscription fees and DSRC\\ networks are not paid\end{tabular} &      \cite{5GMobileCommunications} \\ \hline
\multirow{4}{*}{Network-centric} & Data Traffic & Data delivery ratio and delay &   \cite{When3GMeetsVANET3GAssistedDataDeliveryInVANETs} \\ \cline{2-4} 
 & Load Balancing & Capacities of BSs and RSU &   \cite{ProperHandoverBetweenVANETAndCellularNetworkImprovesInternetAccess} \\ \cline{2-4} 
 & Fairness Guarantees & Between users &   \cite{AFastCloudBasedNetworkSelectionSchemeUsingCoalitionFormationGamesinVehicularNetworks} \\ \cline{2-4} 
 & Network Throughput & Maximize support of user demands &  \cite{NetUtilityBasedNetworkSelectionSchemeInCDMACellularWLANIntegratedNetworks} \\ \hline
\multirow{3}{*}{Hybrid} & \multirow{3}{*}{\begin{tabular}[c]{@{}c@{}}Utility objective based\\ on user- and network-\\ centric handover\end{tabular}} & \multirow{3}{*}{\begin{tabular}[c]{@{}c@{}}Mapped IEEE 1609.4 traffic classes\\ with cellular standards to guarantee \\ expected QoS\end{tabular}} & \multirow{3}{*}{\cite{VerticalMobilityManagementArchitecturesInWirelessNetworksAComprehensiveSurveyAndFutureDirections}} \\
 &  &  &  \\
 &  &  &  \\ \hline
\end{tabular}
\end{table*}
\subsection{Routing Protocols in VANETs}
Finding and maintaining routes in VANETs is a very challenging task especially in networks with very changing topology due to the dynamicity of the vehicular nodes. Similar to IoV, several classical protocols applicable to MANETs are developed, such as: Ad-hoc On Demand Distance Vector (AODV) \cite{AdHocOnDemandDistanceVectorRouting}, Destination-Sequenced Distance-Vector (DSDV) \cite{HighlyDynamicDestinationSequencedDistanceVectorRouting(DSDV)ForMobileComputers} and Dynamic Source Routing (DSR) \cite{DynamicSourceRoutingInAdHocWirelessNetworks}.
Additionally, many research attempts have considered the properties of vehicles and developed geographic protocols like Greedy Perimeter Stateless Routing (GPSR) \cite{GPSRGreedyPerimeterStatelessRoutingForWirelessNetworks} and Greedy Perimeter Coordinator Routing (GPCR) \cite{GeographicRoutingInCityScenarios}.
A detailed taxonomy of existing routing protocols is given in Table \ref{RoutingProtocolsTable}. These protocols are classified in this section based on: 
\begin{itemize}
    \item Required Routing information to perform routing: topology, position, map and path. 
    \item Transmission strategies: geocast, broadcast and unicast. 
    \item Target network type: homogeneous and heterogeneous networks.
\end{itemize}

\subsubsection{Required Routing Information}
The routing decisions are performed by the routing protocols depending on the information provided about the topology of the network, the position of the node, street map details or the path of the packet to be routed from one vehicle to another. In this section, we present the most relevant state of the art VANET routing protocols based on the required information needed to perform packet routing.

\paragraph{Position-Based Routing Protocols}
Position-based routing protocols use the positions of vehicles given by the geographical location information obtained from street maps or on-board Global Positioning System (GPS) systems to make packet forwarding decisions. The GPSR protocol \cite{GPSRGreedyPerimeterStatelessRoutingForWirelessNetworks}, primarily developed for mobile wireless networks, formulates the packet forwarding decisions based on the positions of the nodes. It employs greedy forwarding techniques of packets to vehicles that are progressively closer to the destination. If the greedy path could not be established then the forwarding node forwards the packet based on a planar graph traversal after switching the packet to its perimeter mode. Then it resumes the greedy task when the forwarded packet reaches a node closer to the destination. \\ 
GPSR performance degrades significantly in city scenarios because of obstacles like buildings, which lead to restricted direct communications between nodes. In addition, switching to the planarized graph to find the routing topology and running the greedy algorithm or face routing degrades the performance of packet routing significantly because of excessive delays. \\

Lochert \textit{et al.} \cite{GeographicRoutingInCityScenarios} developed the GPCR protocol that is used to forward packets to vehicles depending on their movements and without the use of source routing or street maps. It considers that vehicles at a junction follow a natural planar graph and packets having first priority are going to be forwarded to a junction node for the sake of determining the next hop. GPCR suffers from several problems like the inability of solving the local maximum problem that happens during forwarding when a vehicle is closer to the destination than its neighbors. In addition, GPCR is very dependent on the node located at the junction to be used used as a coordinator.  \\


In \cite{GeoDTN+NavGeographicDTNRoutingWithNavigatorPredictionForUrbanVehicularEnvironments}, a Geographic Delay Tolerant Network with Navigation (GeoDTN+Nav) routing protocol is developed. It uses a vehicular mobility and the information provided by the on-board vehicular navigation systems to improve the packet delivery for delay tolerant applications in partitioned networks.  GeoDTN+Nav's ability to estimate network partitions and to improve partitions reachability by using a store-carry-forward technique outperforms the packet delivery ratio of GPSR and GPCR. \\

In addition, authors in \cite{ARoutingStrategyForMetropolisVehicularCommunications} developed Anchor-based Street and Traffic Aware Routing (A-STAR) protocol to be deployed in city environments. The protocol computes the chain of a sequence of junctions with traffic awareness until reaching the destination by using statistically rated street maps.   \\
Driscoll \textit{et al.} \cite{HybridGeoRoutingInUrbanVehicularNetworks} proposed a hybrid vehicular routing protocol called Infrastructure Enhanced Geographic Routing Protocol (IEGRP) in a fully deployed RSU environment. IEGRP employs unicast routing and dynamically changes its routing decision under the RSU to maximize the packet delivery rate. \\
Kaiwartya \textit{et al.} proposed a new routing protocol, Traffic light-based Time Stable Geocast (T-TSG) \cite{TrafficLightBasedTimeStableGeocastTTSGRoutingForUrbanVANETs}, for disseminating accident messages in urban vehicular environments by considering the traffic light behavior and vehicle distribution information around an accident. 

\paragraph{Topology-Based Routing Protocols}

Perkins \textit{et al.} \cite{AdHocOnDemandDistanceVectorRouting} developed an AODV routing algorithm primarily applicable to MANETs. It has a route discovery process that broadcasts a route request message (PREQ) and creates many unused routes between source and destination nodes.
In \cite{AnEnhancementToAODVProtocolForEfficientRoutingInVANETAClusterBasedApproach}, the authors developed a cluster based enhanced AODV protocol for efficient routing specifically in VANETs. It improves the performance of the classical AODV protocol by creating stable clusters and performing routing between cluster heads and gateway nodes.  










\vspace{1cm}
\subsubsection{Transmission Strategies}
We classify the routing protocols in VANETs based on the type of information delivery and transmission strategy that they employ; namely, geocast, broadcast and unicast. 

\paragraph{Geocast Routing Protocols}
Geocast routing protocols \cite{ASurveyOfGeocastRoutingProtocols} are location-based multicast algorithms for routing packets from a source vehicle to a multicast group of nodes in a specific geographical zone of relevance.

In \cite{GeocastInVehicularEnvironmentsCachingAndTransmissionRangeControlForImprovedEfficiency}, the authors developed cached greedy geocast and distance aware neighborhood selection schemes to handle unstable routing paths caused by the high mobility of vehicles and the rapidly changing topology of the network. It declares a cache in the routing layer to hold packets temporarily when a node cannot forward packets because of a local minimum and release them from its cache when an affordable close node appears and is selected greedily as the next hop.

Celes \textit{et al.} developed a geocast spatial information (GeoSPIN) \cite{GeoSPINAnApproachForGeocastRoutingBasedOnSPatialINformationInVANETs} based routing protocol for VANETs. The decision making process of message forwarding combines the user trajectories with a geocast strategy in order to improve the data delivery rate in VANETs.

In \cite{AMulticastProtocolInAdHocNetworksInterVehicleGeocast}, the Inter-Vehicles Geocast (IVG) routing protocol is developed to broadcast an alarm message to every vehicle in a specific risk area. 

The T-TSG \cite{TrafficLightBasedTimeStableGeocastTTSGRoutingForUrbanVANETs} is considered as a Geocast routing algorithm that uses the traffic lights' information. When compared to other traditional flooding based-protocols \cite{PerformanceAnalysisOfBroadcastingSchemesInMobileAdHocNetworks}, T-TSG offers higher message delivery rate and lower end-to-end delay.

\paragraph{Broadcast Routing Protocols}
Broadcast is the most used routing technique in VANETs to disseminate traffic and road conditions \cite{UrbanMultiHopBroadcastProtocolForInterVehicleCommunicationSystems}, emergency alerts \cite{EmergencyBroadcastProtocolForInterVehicleCommunications}, advertisements, and infotainment. 

\paragraph{Unicast Routing Protocols}
Unicast based routing protocols in VANETs have the primary purpose of delivering data from one source vehicle or BS to a destination vehicle through multi-hops. The hop-by-hop forwarding mechanism is used to forward data in the quickest manner without making complex forwarding decision \cite{GPSRGreedyPerimeterStatelessRoutingForWirelessNetworks}, \cite{ARoutingStrategyForVehicularAdHocNetworksInCityEnvironments}.  

In the Carry-and-forward forwarding technique, senders and intermediate nodes carry data packets and forward them to a specific node after a forwarding decision is concluded by the underlying routing algorithm \cite{VADDVehicleAssistedDataDeliveryInVehicularAdHocNetworks}.


\subsubsection{Target network}
We classify the common routing algorithms based on the type of source and target networks from/to which we perform routing. Precisely, in the following sub-sections we discuss the classical routing protocols in homogeneous source and destination vehicular networks and we review the existing routing approaches between heterogeneous VANETs. A summary of the main IoV routing protocols and their taxonomy is presented in Table \ref{RoutingProtocolsTable}.

\paragraph{Routing in Homogeneous Vehicular Networks}
Traditional routing protocols in homogeneous vehicular networks assume that packets generated by vehicles are managed via a small distance technology, e.g., WAVE.
Based on \cite{ImprovementOfQoSInVANETWithDifferentMobilityPatterns}, the proposed evaluation performance of homogeneous WiMAX in VANET for many routing protocols DSDV \cite{HighlyDynamicDestinationSequencedDistanceVectorRouting(DSDV)ForMobileComputers}, DSR \cite{DynamicSourceRoutingInAdHocWirelessNetworks} and AODV \cite{AnImprovedAODVRoutingProtocolForVANETsInCityScenarios} validate that the routing protocols perform regardless of the underlying wireless technology.

\paragraph{Routing in Heterogeneous Vehicular Networks}
Heterogeneous routing protocols vary from the homogeneous protocols by including different radio access technologies and consequently routing between heterogeneous networks. Various solutions are developed to switch packets between access technology to another based on vertical handoff \cite{VehicularTelematicsOverHeterogeneousWirelessNetworksASurvey}, nodes clustering and gateway selection \cite{AnEfficientQosBasedGatewaySelectionAlgorithmForVANETToLTEAdvancedHybridCellularNetwork} and routing protocols from 3G to different RSU access \cite{When3GMeetsVANET3GAssistedDataDeliveryInVANETs} and between WLAN and WiMAX hops double technology routing \cite{WLANWiMAXDoubleTechnologyRoutingForVehicularNetworks}, etc.

\begin{table*}[t]
\centering
\caption{TAXONOMY OF ROUTING PROTOCOLS IN IOV}\label{RoutingProtocolsTable}
\begin{tabular}{|c|c|c|c|c|l|}
\hline
\multicolumn{1}{|l|}{Routing Protocols} & Routing Category & \multicolumn{1}{l|}{Routing Type} & Delay Sensitivity & Information Used & \multicolumn{1}{c|}{Reference} \\ \hline
AODV & Topology-based Routing & Unicast & Delay-sensitive & Topology-based & \cite{AdHocOnDemandDistanceVectorRouting}, \cite{AnEnhancementToAODVProtocolForEfficientRoutingInVANETAClusterBasedApproach} and \cite{AnImprovedAODVRoutingProtocolForVANETsInCityScenarios}\\ \hline

AODV-Bis & Topology-based Routing & Unicast & Delay-sensitive & Route-Req-Forwarding & \cite{ImplementationOfGeocastEnhancedAODVBisRoutingProtocolInMANET} \\ \hline
PRAODV-M & Topology-based Routing & Unicast & Delay-sensitive & Route-Selection & \cite{AStudyOnTheFeasibilityOfMobileGatewaysForVehicularAdHocNetworks} \\ \hline
DSR & Topology-based Routing & Unicast & Delay-sensitive & Topology-based & \cite{DynamicSourceRoutingInAdHocWirelessNetworks} \\ \hline
DSDV & Topology-based Routing & Unicast & Delay-sensitive & Topology-based & \cite{HighlyDynamicDestinationSequencedDistanceVectorRouting(DSDV)ForMobileComputers} \\ \hline
Flooding UMB & Broadcast Routing & Broadcast & Delay-sensitive & Packet Forwarding & \cite{UrbanMultiHopBroadcastProtocolForInterVehicleCommunicationSystems} \\ \hline
BROADCOMM & Broadcast Routing & Broadcast & Delay-sensitive & Formation of Cells & \cite{EmergencyBroadcastProtocolForInterVehicleCommunications} \\ \hline

GPSR & Position-Based Routing & Unicast & Delay-tolerant & Packet Forwarding & \cite{GPSRGreedyPerimeterStatelessRoutingForWirelessNetworks}, \cite{AGPSREnhancementMechanismForRoutingInVANETs} \\ \hline

GPCR, GpsrJ+ & Position-Based Routing & Unicast & Delay-tolerant & Packet Forwarding & \cite{GeographicRoutingInCityScenarios}, \cite{EnhancedPerimeterRoutingForGeographicForwardingProtocolsInUrbanVehicularScenarios} \\ \hline
GeoDTN+Nav & Position-Based Routing & Unicast & Delay-tolerant &  Store-Carry-Forward  & \cite{GeoDTN+NavGeographicDTNRoutingWithNavigatorPredictionForUrbanVehicularEnvironments} \\ \hline

A-STAR & Position-Based Routing & Unicast & Delay-tolerant & Packet Forwarding and Traffic Info & \cite{ARoutingStrategyForMetropolisVehicularCommunications} \\ \hline
IEGRP & Position-Based Routing & Unicast & Delay-tolerant & Packet Forwarding & \cite{HybridGeoRoutingInUrbanVehicularNetworks} \\ \hline
IVG & Position-Based Routing & Geocast & Delay-sensitive & Packet Forwarding & \cite{AMulticastProtocolInAdHocNetworksInterVehicleGeocast} \\ \hline
T-TSG & Position-Based Routing & Geocast & Delay-sensitive & Packet Forwarding & \cite{TrafficLightBasedTimeStableGeocastTTSGRoutingForUrbanVANETs} \\ \hline
Cached geocast & Position-Based Routing & Geocast & Delay-tolerant & Packet Forwarding & \cite{GeocastInVehicularEnvironmentsCachingAndTransmissionRangeControlForImprovedEfficiency} \\ \hline
GeoSpin & Position-BasedRouting & Geocast & Delay-tolerant & Packet Forwarding & \cite{GeoSPINAnApproachForGeocastRoutingBasedOnSPatialINformationInVANETs} \\ \hline
GeoSVR & Map-Based Routing & Unicast & Delay-tolerant & Packet Forwarding & \cite{GeoSVRAGeographicStatelessVANETRouting} \\ \hline
GSR & Map-Based Routing & Unicast & Delay-tolerant & Packet Forwarding & \cite{ARoutingStrategyForVehicularAdHocNetworksInCityEnvironments} \\ \hline
STAR & Map-Based Routing & Unicast & Delay-tolerant & Packet Forwarding & \cite{IntersectionBasedRoutingForUrbanVehicularCommunicationsWithTrafficLightConsiderations} \\ \hline
VADD & Path-based & Unicast & Delay-tolerant & Packet Forwarding & \cite{VADDVehicleAssistedDataDeliveryInVehicularAdHocNetworks} \\ \hline
\end{tabular}
\end{table*}
\section{Traffic modeling and networking of Safety and Non-Safety Applications}
\label{SectionTrafficModelingandNetworkingofSafetyandNon-SafetyApplications}
To better understand how the traffic density in vehicular networks affects the reliability of the safety applications in event-driven and periodic applications, many studies \cite{OptimizingtheControlChannelIntervaloftheDSRCforVehicularSafetyApplications} allow safety applications to optimize the CCHI while applying different types of workloads composed of various safety applications. Many types of safety applications depend on different types of data  \cite{TrafficModelingofSafetyApplicationsInVehicularNetworks}.\\
In the VANET WAVE Protocol, every CCH period supports sending only one beacon packet for every vehicle in every channel period. The key point is how to prevent and detect the collision caused by hidden or exposed terminals in the broadcast mode especially that there are no Request To Send (RTS), Clear To Send (CTS), Acknowledgements (ACKs) and re-transmission packets as in the case of IEEE 802.11 wireless networking protocols. There are many types of safety applications that are used to increase the safety of roads by exchanging information regarding the traffic, vehicles' statuses, road conditions and many other details. These applications include: 

\begin{itemize}
    \item V2I safety applications \cite{VehicletoInfrastructureV2ISafetyApplications} related to the 
 communications between vehicles and road infrastructure. They include Stop Sign Violation Warning (SSVW) \cite{pathan2016security}, Railroad Crossing Violation Warning (RCVW), Spot Weather Information Warning (SWIW), Oversize Vehicle Warning (OVW) and Reduced Speed Zone Warning (RSZW). 
    \item V2V safety applications \cite{VehicleSafetyCommunicationsApplicationsVSCA} that are executed by vehicles and received by their neighboring ones. They include Forward Collision Warning (FCW), Electronic Emergency Brake Light, Do Not Pass Warning, Left Turn Assist (LTA), Intersection Movement Assist (IMA), Blind Spot Warning (BSW) and Lane Change Warning (LCW). 
\end{itemize} 
The authors in \cite{TrafficModelingofSafetyApplicationsInVehicularNetworks} studied the packet reception rate of various scenarios each composed of a mixture of different safety applications. Packet reception rates in VANETs characterizes the successful transfer probability of safety packets sent between vehicles and it is calculated to achieve the overall successful probability of transfer. The probability is based on the following three conditions: successful transfer probability of a packet when no other vehicle is sending, the successful transfer probability of a packet when left side senders are sending packets, and the successful transfer probability of a packet when right side senders are sending packets. \\

Asgari \textit{et al.} \cite{TrafficModelingofSafetyApplicationsInVehicularNetworks} proposed that Based on the Poisson Distribution rules, the workload of a running application on a vehicle can be characterized by one parameter \textit{$\lambda${r}} which is the exponential distribution parameter of the time between packets received by other vehicles and time of the presented mathematical model. A very challenging problem is to allow vehicles to access the channel with a derived optimal probability in a way that maximizes the successful transmission rates. It also modifies DSRC parameters based on the conditions of the road and network conditions to allow the co-existence of safety and non-safety applications with lower congestion and higher successful transmission rate. \\

In \cite{AFastCloudBasedNetworkSelectionSchemeUsingCoalitionFormationGamesinVehicularNetworks}, authors presented cloud-based network selection scheme to ensure that vehicles on the move are assisted and are able to make decisions to select the best network after having a wider network awareness scope. The developed scheme considers the high mobility of vehicles and the fast changing network topology as well as the changes adapted to the communications channel leading to unstable Received Signal Strength (RSS). Xu \textit{et al.} \cite{AFastCloudBasedNetworkSelectionSchemeUsingCoalitionFormationGamesinVehicularNetworks} developed a coalition formation game model for vehicular networks with a fast convergence algorithm tested with three vehicular scenarios that demonstrate a realistic range of network topology, resource availability and mobility patterns of vehicles. It uses a distributed database that holds and periodically updates the networks' statistics, road traffic flow information and network maps. Vehicles can consequently apply a network selection algorithm to determine the most suitable network that maximizes their flow of application packets from and to the cloud. 
\section{Vehicular Cloud Computing (VCC)}\label{SectionVehicularCloudComputing}
Realizing a standardized VCC is not an easy task to achieve due to the various proposed architectures and the many challenges related to various provenances of computing and virtualization capabilities.
\subsection{Vehicular Cloud (VC) Architecture}
Based on the description in \cite{ASurveyOnVehicularCloudComputing} and in \cite{VehicularCloudNetworksArchitectureApplicationsAndSecurityIssues}, the overall VCC architecture relies on three layers: 
\begin{itemize}
    \item Inside-vehicle or Tier-1 cloud formed by physical resources.
    \item Vehicular and infrastructure communications or Tier-2 cloud.
    \item The abstraction of computational resources and the formation of VCC similar to a traditional Back-End Cloud (BEC).
\end{itemize}
The inside-vehicle layer \cite{HealthDriveMobileHealthcareOnboardVehiclesToPromoteSafeDriving} is responsible for monitoring the health and mood of the driver and collecting information inside the vehicle.  The collected data can be pressure and temperature by using body sensors, environmental sensors, smart phone sensors, the vehicle's internal sensors, Inertial Navigation Sensors (INS), and driver behavior recognition to predict the driver's reflexes and intentions. \\
Later, all the information collated via sensors should be sent to the cloud for storage or for use as input for various software programs in the application layer. It is assumed that each vehicle is equipped with an OBU and that is considered as the key-stone for computation in the VC. Equally important, the OBUs have broadband wireless communications capabilities to transfer data through 3G or LTE cellular communications devices, Wi-Fi, WiMAX, or WAVE. The OBU is connected to the sensors, GPU, GPS, controller, etc. through a central gateway and form a complex in-vehicle network \cite{WirelessSensorAndMobileAdHocNetworksVehicularAndSpaceApplications}.
The next layer of this architecture is responsible for the communications. This layer covers V2V and V2I communications. The V2I component of this layer is used to augment the safety level of vehicles on highways by reducing the percentage of crashes, delays and congestion, improve mobility, and provide Wireless Roadside Inspection (WRI) to automatically inspect commercial vehicles. The cloud is the last layer of the VCC architecture that allows for massive and complex computations in minimal time. The cloud layer consists of three internal sub-layers: application, cloud infrastructure, and cloud platform. \\
\subsection{Vehicular Cloud Computing (VCC) Challenges}
\par The computing resources of vehicles that remain underutilized are combined to create a vehicular cloud composed of vehicles in parking lots \cite{TowardsFaultTolerantJobAssignmentInVehicularCloudComputing}. In \cite{RSUCloudAndItsResourceManagementInSupportOfEnhancedVehicularApplications},
the authors propose forming a vehicular cloud by using the microdatacenters in the RSUs so that every RSU offers its capabilities to run infotainment applications. In \cite{CostMinimizationSchedulingforDeadlineConstrainedApplicationsOnVehicularCloudInfrastructure} the VC is able to merge autonomously the moving OBUs to offer services. 
\begin{figure}[h]
\includegraphics[width=9cm,height=16cm,keepaspectratio]{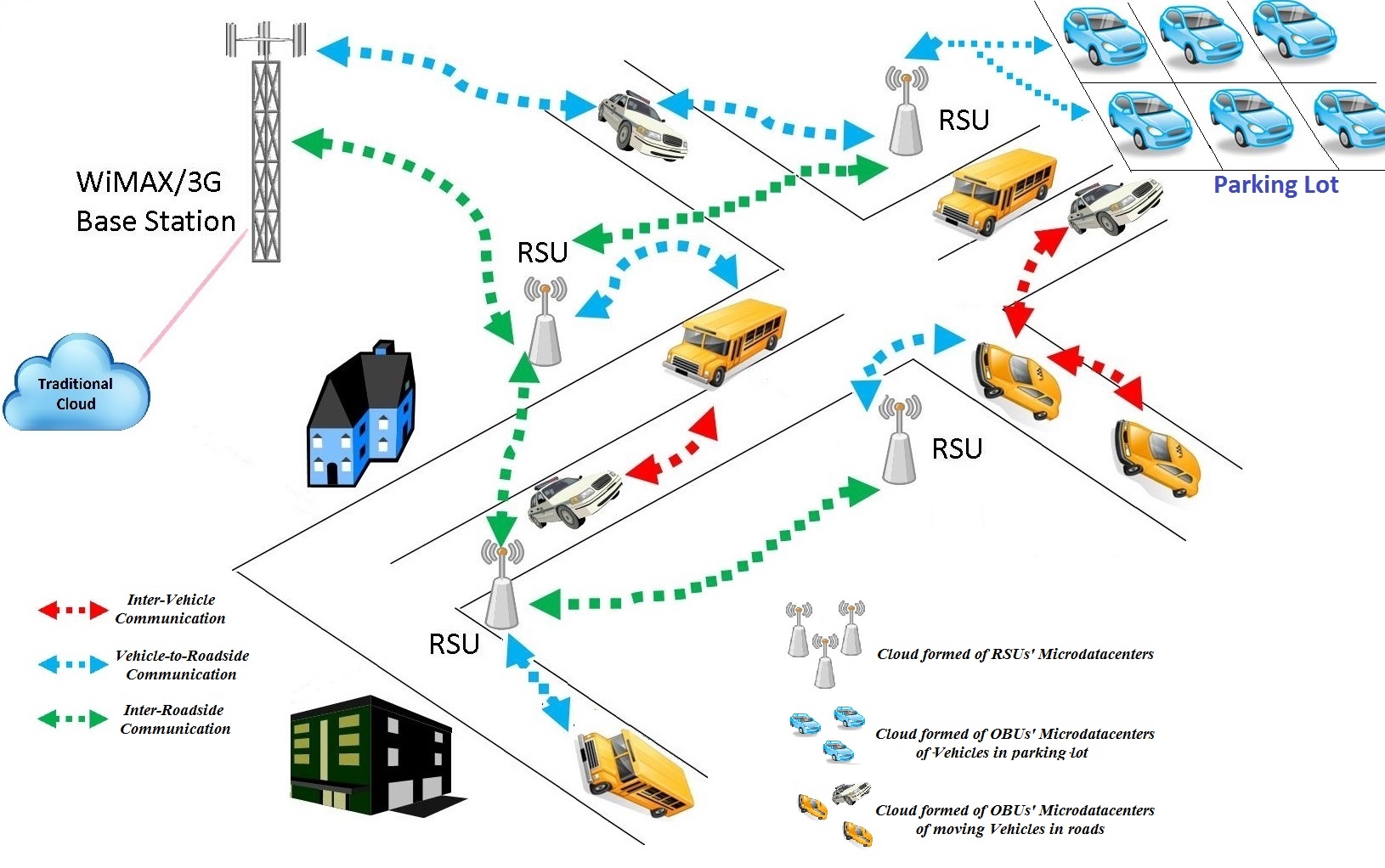}
\centering
\caption{Different Architectures of Vehicular Clouds.}
\end{figure}
Whether the VC is static or in motion, many challenges still remain on how to manage these types of clouds with unpredictable availability of computational resources as vehicles enter and leave the cloud. Moreover, handling VMMs with the lowest cost possible and across heterogeneous VCs is a requirement. The technical challenge also lies in the fact that the vehicles come and go, which makes the assignment of computing resources to tasks difficult. In addition, there is a tremendous need for fault tolerant job assignment strategies to mitigate the effects of resource volatility and unavailability in VCs, as detailed in \cite{TowardsFaultTolerantJobAssignmentInVehicularCloudComputing}.\\ This can be achieved by offering a theoretical analysis and prediction of the expected job completion time in both cases when vehicles do not leave during a checkpoint operation and when vehicles leave while checkpointing is in progress, which leads to a loss of VMs context and to the failure of the system. One of the solutions proposed in the literature is referred to as checkpointing where the state of the computation is saved periodically while jobs are still in progress and the most recent image, or also called copy, of VM is used to rollback the computation in case of a system failure. The contribution of the authors in this paper is focused on saving the state of the computation of a VMM by a vehicle only when needed.\\ This puts less overhead on the system, especially that most of the checkpoints are not necessary which may increase the efficiency of the vehicular cloud. Ghazizadeh \textit{et al.} \cite{TowardsFaultTolerantJobAssignmentInVehicularCloudComputing} designed a model for deterministic environment in terms of vehicle departures and arrivals to the network. The authors explored also the idea of assigning each job to two vehicles for fault-tolerance purposes and taking a checkpoint only when one of them leaves the parking lot.\\ The proposed solution seems efficient in terms of minimizing the packet losses of applications when vehicles leave the cloud but it has a significant disadvantage as each task is being executed twice at the same time which decreases the effective throughput of the cloud. 
\par In \cite{CooperativeResourceManagementInCloud-EnabledVehicularNetworks} the authors investigate the resource management and sharing problem for bandwidth communication and computing resources to handle mobile applications in cloud-enabled vehicular networks. Cloud SPs cooperate to form coalitions of cloud resources to share their idle resources with each other. The coalition game model or scheme based on two-sided matching theory for cooperation among cloud SPs. The resources can be better utilized with an improved QoS for users.\\
The Cloud Service Provider (CSP) reserves a certain amount of long-term bandwidth from network providers and reserves some long-term computing resources (eg., CPU, Memory, Storage, Bandwidth, etc.) from data-centers, which are owned by CSPs. When the CSP’s available resources are limited and do not satisfy the vehicular application requirements, the service quality and user experience degrade. With this in mind, the solution is that vehicles share their storage, bandwidth, and energy resources with each other to secure, store, carry and forward network data for the mutual enhancement of the overall performance of a network of the cloud environment.\\
The Vehicular Cloud Service Providers (VCSPs) using the computation and communications capabilities of the underlying OBUs have to employ efficient resource management and sharing schemes to meet the requirements of the applications on the street, while taking into account the main differences between Mobile Cloud Computing (MCC) and VCs especially the high mobility of vehicles. It is considered as a compelling business model in the cloud market to know the cost of resources to be rented and the reward of resources offered to the cloud. Having a combination of prices with resource demands in cloud-enabled vehicular networks is implemented as a coalition formation application using the two sided matching theory.\\
This application also helps to improve the fairness of transactions and optimizes the resource utilization and cost of the VCSP. Therefore, building a win-win situation by rewarding, as in \cite{VehicularCloudNetworkingArchitectureAndDesignprinciples}, vehicles agreeing to become a cloud member.
\par Additionally, some applications have strict deadline constraints, which causes a problem with decision-making for VC selection to host the applications especially with unstable VC resources. In  \cite{CostMinimizationSchedulingforDeadlineConstrainedApplicationsOnVehicularCloudInfrastructure},  two major issues that are considered related to the cloud availability and lifetime, and the applications processing time. The high mobility of vehicles over time reflects the dynamic changes of resource availability over time, which similarly implies many issues regarding the management of the the dynamic VC's resources to guarantee that tasks are completed with minimum cost before their deadlines and within the lifetime of the VC. In contrast to the previously discussed work, authors in \cite{CostMinimizationSchedulingforDeadlineConstrainedApplicationsOnVehicularCloudInfrastructure} assume that the  computing resources of the OBUs are limited and could merge autonomously, flexibly and dynamically to offer services to authorized users. The proposed architecture of the dynamic VC consists of a Cloud Directory Application Repository (CDAR) that contains a list of all applications, all tasks that compose applications and application deadlines, vehicular cloud communications (e.g., WAVE, 4G or 5G Based).\\
In addition, the architecture is based on a broker that first announces invitations for participation in the cloud, receives responses of the vehicles containing the type and the amount of the available resources and finally registers the vehicular cloud in a CDAR in its area and specifies the properties of resources, lifetime of cloud and costs of resources. 


In addition, one major concern in VC is the cloud formation and VMM specifically with the most challenging type of VC that has mobile micro datacenters mounted on vehicles' OBUs and with RSUs acting as head of cloud. In \cite{DynamicVirtualMachineMigrationInaVehicularCloud}, authors have developed dynamic VMM for dynamic VCs. When a vehicle is moving it may exit one network and enter another. If it is still in the same area or still in the communications range of the same RSU, there is no need for VMM. Otherwise, there will be a pressing need to effectively migrate the virtual machines running those applications in order to keep them in the same network. The authors proposed a new scheme called vehicular VMM which is used to better achieve an effective handling of frequent changes in the data center topology and host heterogeneity while keeping minimal RSU intervention.

\begin{table*}[t]
\begin{center}
\caption{VEHICULAR CLOUD COMPUTING SERVICES}
\begin{tabular}{|c|c|c|l|}
\hline
Cloud Architecture                                                                          & Description                                                                                                                                                               & Characteristics                                                                                                                              & \multicolumn{1}{c|}{Reference} \\ \hline
\multirow{2}{*}{\begin{tabular}[c]{@{}c@{}}Network as a \\ Service (NaaS)\end{tabular}}     & \multirow{2}{*}{\begin{tabular}[c]{@{}c@{}}Vehicles offer their Internet \\ connection while on the move\end{tabular}}                                                    & VNAODV                                                                                                                                       &\cite{VirtualizationInVANETsToSupportTheVehicularCloudExperimentsWithTheNetworkAsAServiceModel}                                \\ \cline{3-4} 
                                                                                            &                                                                                                                                                                           & \begin{tabular}[c]{@{}c@{}}Distributed Vehicles \\ without RSU\end{tabular}                                                                  &\cite{CooperationAsAServiceInVANETs}                                \\ \hline
\multirow{3}{*}{\begin{tabular}[c]{@{}c@{}}Storage as a \\ Service\\  (STaaS)\end{tabular}} & \multirow{3}{*}{\begin{tabular}[c]{@{}c@{}}Vehicles offer their extra\\  storage capabilities\end{tabular}}                                                               & \begin{tabular}[c]{@{}c@{}}Backup purposes or for   \\ p2p  applications\end{tabular}                                                        &\cite{TheNextParadigmShiftFromVehicularNetworksToVehicularClouds}                                \\ \cline{3-4} 
                                                                                            &                                                                                                                                                                           & Replication-based storage                                                                                                                    &\cite{RemoteDataCheckingForNetworkCodingBasedDistributedStorageSystems}                                \\ \cline{3-4} 
                                                                                            &                                                                                                                                                                           & \begin{tabular}[c]{@{}c@{}}Vehicles in parking area and\\  RSU for storage management\end{tabular}                                           &\cite{ImplementingStorageAsAServiceInVANETUsingCloudEnvironment}                               \\ \hline
\multirow{2}{*}{\begin{tabular}[c]{@{}c@{}}Cooperation as\\ a Service (CaaS)\end{tabular}}  & \multirow{2}{*}{\begin{tabular}[c]{@{}c@{}}Clustered Vehicles give \\ important information about \\ service for subscribers\end{tabular}}                                & \begin{tabular}[c]{@{}c@{}}Content-Based Routing for \\ intra cluster communications\end{tabular}                                            &\cite{DatacenterAtTheAirportReasoningAboutTimeDependentParkingLotOccupancy}                                \\ \cline{3-4} 
                                                                                            &                                                                                                                                                                           & \begin{tabular}[c]{@{}c@{}}Provides free services\\  without additional infrastructure\end{tabular}                                          &\cite{CooperationAsAServiceInVANETs}                                \\ \hline
\begin{tabular}[c]{@{}c@{}}Information \\ as a Service\\  (INaaS)\end{tabular}              & \begin{tabular}[c]{@{}c@{}}Information that increase safe \\ driving like advance warning \\ of emergency situations, road \\ conditions, news of events etc\end{tabular} & \begin{tabular}[c]{@{}c@{}}Services result of \\ vehicles cooperation  that\\  include all the factors \\ improving the driving\end{tabular} &\cite{CooperationAsAServiceInVANETs}, \cite{TIaaSSecureCloudAssistedTrafficInformationDisseminationInVehicularAdhocNETworks}                                \\ \hline
\begin{tabular}[c]{@{}c@{}}Software as a \\ Service (SaaS)\end{tabular}                     & \multicolumn{1}{l|}{\begin{tabular}[c]{@{}l@{}}Entertainment and infotainement\\  applications  and any  kind of \\ software running ad hoc clouds.\end{tabular}}         & \begin{tabular}[c]{@{}c@{}}Deployed between mobile \\ vehicles in VANET\end{tabular}                                                         &\cite{ANewSoftwareBasedServiceProvisionApproachForVehicularCloud}                                \\ \hline
\multirow{2}{*}{\begin{tabular}[c]{@{}c@{}}Platform as a \\ Service (PaaS)\end{tabular}}    & \multirow{2}{*}{\begin{tabular}[c]{@{}c@{}}Vehicles resources and \\ services are being accessed by a code\end{tabular}}                                          & PaaS for High Scale                                                                                                                          &    \cite{ACloudPAASforHighScaleFunctionandVelocityMobileApplicationsWithReferenceApplicationAsTheFullyConnectedCar}                            \\ \cline{3-4} 
                                                                                            &                                                                                                                                                                           & Cloud4SOA                                                                                                                                    &    \cite{CloudPaaSBrokeringinActionTheCloud4SOAManagementInfrastructure}                            \\ \hline
\end{tabular}
\end{center}
\end{table*}
\subsection{Software Defined Wireless Networking, Fast Greedy Heuristic and CUDA Accelerated with Optimal Scheduling of Infotainment Tasks Placement Over VCs}
\label{SectionSoftwareDefinedWirelessNetworking}
In \cite{SoftwareDefinedNetworkingForRSUCloudsInSupportOfTheInternetOfVehicles}, the authors
developed the idea to interchangeably use VM migration and service migration in an RSU-based cloud. In order to adapt the network with changing services demand, the authors studied the increase or decrease in the number of micro datacenters by having enough resource hosting services in every location  and physically migrating the VMs hosting the services from one micro datecenter to another via the data plane. The authors had to deal with issues related to the limited bandwidth, network link latency,  forwarding information cost, and reconfiguration overhead. The proposed RSU-based cloud CRM with SDN is used to select the best configuration that minimizes the reconfiguration overhead, which minimizes the cost of reconfiguration of service hosting, service migration, service replication and data forwarding rules in the SDN of the RSUs (with micro datacenters hosting non-safety services). The cloud manager communicates via the data-plane with OpenFlow and Cloud Controllers (CCs) to disseminate information regarding service hosting, service migration, data flow changement. In the data plane, the CCs handle service migration and hypervisors to instantiate or delete new VM hosting services and later the OpneFlow controllers update the switch flow rules via the control plane.
The VC can be tested using real-life benchmark programs through some new performance metrics and precisely making the benchmark experiments for the IaaS cloud platforms scale-out (or horizontal scaling which means spreading out the system and adding more Virtual Machines (VMs) from vehicles beside the existing ones) and scale-up (or vertical scaling by replacing one vehicle by a more capable one that is logically equivalent) to cope with workloads. Such cloud management leads to prompt scheduling and placement with awareness of additional costs due to VM migration, network latency and cost.

As opposed to \cite{SoftwareDefinedNetworkingForRSUCloudsInSupportOfTheInternetOfVehicles} where the placement of tasks is made based on minimizing VM migration over demand changes, a Fast greedy heuristic algorithm is presented in \cite{VanetCloudCudaAccelerated} in order to find sub-optimal solutions for virtual
machine allocation and vehicular task placement based on tasks' requirements and VCs' availability, in terms of free VMs. The greedy scheduling algorithm does not reconsider its placements after a scheduling decision of tasks is made. Consequently, it will result in non-optimal placement decisions of vehicular tasks over VCs because one policy may have the current highest reward for task placement among all the current possible decisions, but may deteriorate the reward of the VCs after next placement decision is made.

The Markov Decision Process (MDP) based vehicular task scheduling over VCs based on the opportunistically available V2I communications between VCs is presented in \cite{VanetCloudCudaAccelerated}. The MDP-based task scheduling in VCC is developed to optimize the task placement of various vehicular tasks over
many VCs in a way that increases the maximum expected long-term cumulative reward of the VCC and not for minimizing the cost of VM migration from one VC to another. Then, the authors presented and accelerated version of the MDP scheme that uses the Block of State Divided Iteration (BSDI) algorithm to parallelize the state space of the VCC and to adapt the value iteration algorithm to run in parallel using the CUDA programming
model.
The challenges of vehicular task placement and scheduling in VCCs are mainly constrained by the abundance of VMs, the proximity of application to the users' (running in a VC near the vehicle that requests the application), the optimality of placement in terms of VM utilization and minimization of resource leasing costs from traditional clouds.
\section{Security threats and attacks in Vehicular communication and cloud }\label{SectionSecurityThreats}
With its highly dynamic topology, VANETs are posing a real security threat that has to be addressed, especially with the V2V broadcast communications. Many attacks may occur with different damage levels and threatening vehicles safety on roads. VCs have many vulnerabilities as well, which limits their expansion in the real world, mainly related to exposing sensitive personal data from the cloud to other vehicles. In this section, we go through the challenges facing the IoV and describe some of the recent research solutions. 
\subsection{Threats and Vulnerabilities in Vehicular Communication}
Major attacks, threats and vulnerabilities can be devastating to vehicular communications and to the VANETs safety packets that may contain life critical information. Most common threats studied in the literature include but not limited to:
\begin{itemize}
    \item Vehicles intentionally generating packets to Jam other communications \cite{raya2007securing}. 
    \item Vehicles snooping all the communications of surrounding vehicles or refusing to transmit messages received from other vehicles in the same network \cite{5989903}.
    \item Vehicles using misleading messages to make vehicles avoid a fictional congestion ahead and free the road, escape from police pursuit by creating blocking vehicles, disrupting neighboring vehicles with imaginary risky situations or by creating an imaginary event \cite{papadimitratos2008secure}.   
\end{itemize}

\subsection{Threats and Attacks in Vehicular Clouds (VCs)}
In order have a wide adoption, VC should overcome many security issues and privacy threats. In addition to the security challenges in cloud computing, specific VC vulnerabilities need to be studied concerning the trust relationship between vehicles, identity tracking of high-mobility vehicles, and coalition formation of VC. Detailed potential issues in vehicular clouds are studied in \cite{SecurityChallengesInVehicular}. In the following paragraphs, we present the most prominent challenges:
\begin{itemize}
    \item Denial of Service (DoS) \cite{hasbullah2010denial} attacks created mainly by the generation of jamming transmission or recurrent non needed requests interfering with other true messages and leading to a total failure by preventing successful transmission of packets to clouds. 
    \item Hacker vehicles entering a coalition of vehicles forming a cloud to spoof and steal the identity of other vehicles and to retrieve sensitive data originating from the cloud \cite{SecurityChallengesInVehicular}. 
    \item Malware software that changes the contents of secure data stored in the cloud representing traffic patterns or accidents. \cite{SecurityChallengesInVehicular}.
\end{itemize}

\begin{figure}[h] 
\begin{center} 
\includegraphics[width=9cm,height=12cm,keepaspectratio]{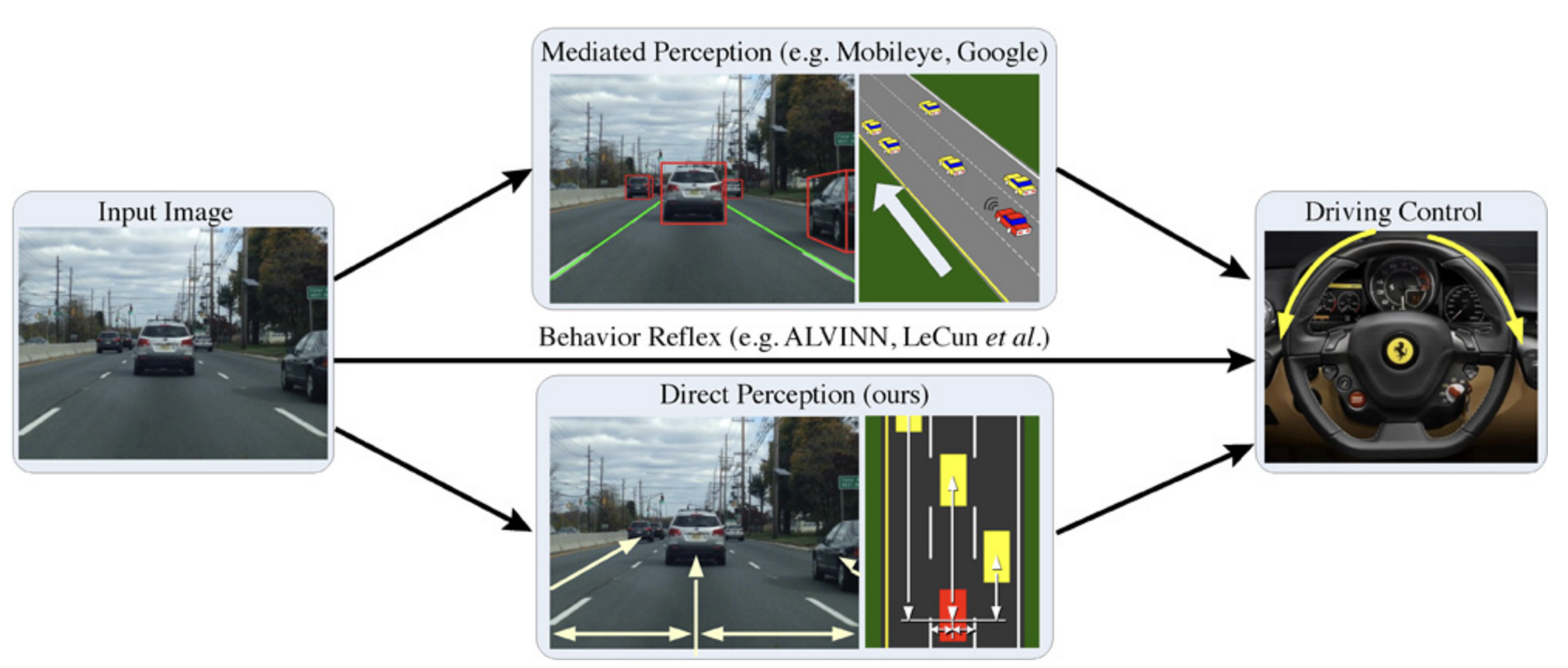}
\end{center}
\caption{Illustration of Three Paradigms of the Autonomous Driving Systems \cite{DeepDrivingLearningAffordanceForDirectPerceptionInAutonomousDriving}.}
\label{IllustrationOfThreeParadigmsOfTheautonomous}
\end{figure}

\textbf{ {\normalsize PART 3: AUTONOMOUS DRIVING: PERCEPTION AND LIDAR SCENE UNDERSTANDING AND RECONSTRUCTION} } 
In this section we go through the most successful machine learning techniques for AD approaches using different input modality sensors, as presented in Fig. \ref{IllustrationOfThreeParadigmsOfTheautonomous}. With this in mind, the classification of recent work falls into the following broad categories: mediated direct perception approaches, behavior approaches, reflex approaches, and LIDAR perception approaches. Then, we present recent visual-based and LIDAR-basd odometry for vehicle localization and mapping based on raw data with non-recognized objects in scene, based on camera-only feeds, LIDAR-only point cloud data, and mixed approaches combining visual and LIDAR inputs for joint vehicle localization and mapping within its surrounding objects.

\section{Fully Autonomous Self Driving Vehicles: Machine Learning as an Alternative Approach to Computer Vision}\label{SectionSecurityThreats}

\subsection{Mediated Perception Approaches for Autonomous Driving Systems}
Mediated perception approaches process the entire scene given by a front-facing camera and parse it in order to recognize driving-relevant objects that impact the driving decisions. This kind of approach depends on independent modules that are used for different specialized recognition tasks, such as lanes, vehicles, pedestrians, traffic lights and signs, free space, etc. \cite{VisionMeetsRoboticsTheKITTIDataset}. \\
The recognition results of these modules are integrated and combined in a unique world representation to represent the surroundings of the vehicle and from which the final driving decisions are made, as detailed in most of the state-of-the-art systems \cite{AutonomousGroundVehicles}. The complexity and cost of mediated perception approaches stems from the need to make a total scene parsing and understanding, while only a small number of recognized objects are relevant to make driving decisions from one frame to another. 
\subsubsection{Pixel by Pixel Full Scene Labeling (FSL)}
\begin{figure}[h] 
\begin{center} 
\includegraphics[width=\columnwidth]{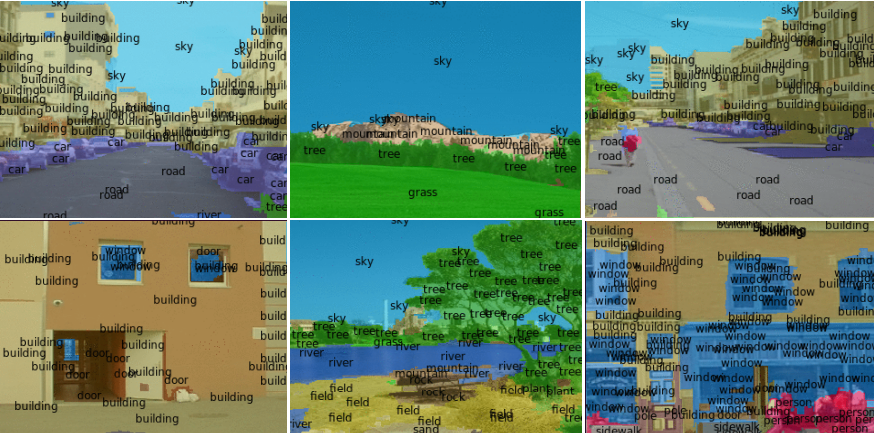}
\end{center}
\caption{Results of FSL on SIFT Flow Dataset \cite{LearningHierarchicalFeaturesForSceneLabeling}. } 
\label{ResultsOfFSLonSIFTFlowDataset}

\end{figure}

One approach to understand the scene around vehicles is semantic segmentation that labels each pixel in an image with the category of the object to which it belongs. Labeling each pixel of the scene independently from its surrounding pixels is a very hard task to achieve. In order to know the category of a pixel, the labeling process relies on relatively short-range surrounding information and long-range information to understand the context while an object covers more than a pixel. Figure \ref{ResultsOfFSLonSIFTFlowDataset} shows the result of FSL on a camera frame from the SIFT flow Dataset. \\
In other words, to determine that a certain pixel belongs to a vehicle, pedestrian or to any other class of object, FSL needs to have a contextual window that is wide enough to show the surrounding of the pixel and consequently to make an informed decision of the class of the object that contains the pixel. Techniques based on Markov Random Fields (MRF), Conditional Random Field (CRF) and other graphical models are presented in 
\cite{LearningHybridModelsForImageAnnotationwithPartiallyLabeledData}, 
\cite{AssociativeHierarchicalCRFsForObjectClassImageSegmentation}, \cite{StackedHierarchicalLabeling} to guarantee the consistency of labeling the pixels in the context of the overall image. In addition, the authors in \cite{SuperparsingScalableNonparametricImageParsingWithSuperpixels}, \cite{EfficientlySelectingRegionsForSceneUnderstanding} and \cite{DecomposingASceneIntoGeometricAndSemanticallyConsistentRegions} developed various methods for pre-segmentation into super-pixels or segment candidates that are used to extract the categories and features characterizing individual segments and combinations of neighboring segments.

Farabet \textit{et al.} \cite{SceneParsingWithMultiscaleFeatureLearningPurityTreesAndOptimalCovers} developed a FSL technique that includes detection, segmentation and recognition of all objects in the scene. It uses a large contextual window that labels pixels and reduces the requirements for postprocessing methods by using a multiscale Convolutional Neural Network (CNN) that is trained from raw pixels to extract dense feature vectors that are encoding regions of multiple sizes centered on each pixel of the image.\\

A complete detailed version of the same scene parsing architecture is developed in \cite{LearningHierarchicalFeaturesForSceneLabeling}. The key idea behind the proposed mutiscale dense feature extractor is to generate a series of feature vectors of regions with multiple sizes and centered around every pixel. The CNNs that are fed with raw pixels with end to end training have copies of a single network with the same weights. These networks are applied to multiple scales of a Laplacien pyramid version of the input image. The trained networks offer features that produce efficient multiscale representations for FSL to capture texture, shape and contextual information. Even though the learned multiscale representations allow for the detection and recognition of regions and objects contained in the scene, they do not draw the boundaries of the regions accurately.

\begin{figure}[h] 
\begin{center} 
\includegraphics[width=\columnwidth]{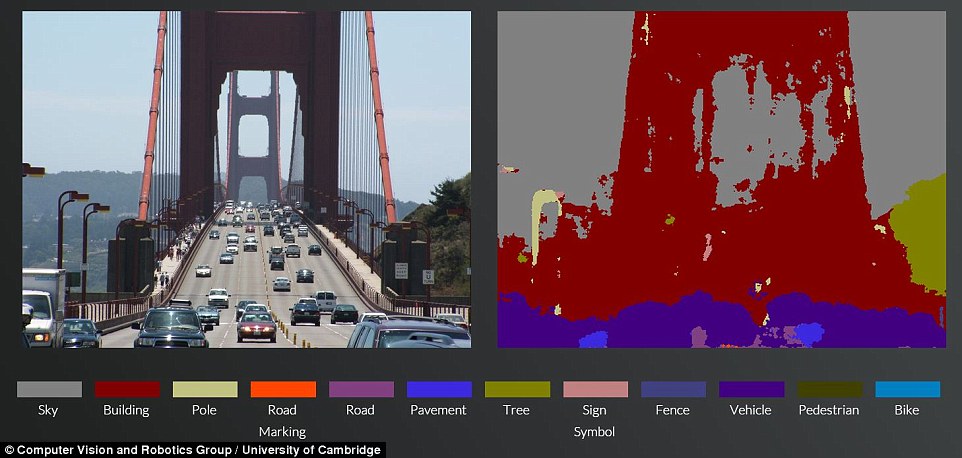}
\end{center}
\caption{Pixels Labeling Based on Different Class of Objects \cite{SegNetADeepConvolutionalEncoderDecoderArchitectureforImageSegmentation}.} 
\label{PixelsLabelingBasedOnDifferentClassOfObjects}
\end{figure}

In \cite{SegNetADeepConvolutionalEncoderDecoderArchitectureforImageSegmentation}, the authors were the first to utilize the CNN architecture for semantically segmenting the pixel of the images using a pixel-wise classification layer on top of the the decoder network that follows the encoder network. Impressive results of the pixe-lwise classification are presented in Fig. \ref{PixelsLabelingBasedOnDifferentClassOfObjects}.

\subsubsection{3D Scene Flow Segmentation}
The scene flow is considered as a promising technique in understanding the scene for autonomous vehicles. It offers a flow field describing the 3D motion of interesting objects in the scene by jointly creating their dense geometry and 3D motion from sequences of images. A real-time 3D reconstruction of the scene is eventually achievable by associating every pixel and its exact 3D position with the significant object to which it belongs. It would require solving the open problem of monocular 3D reconstruction of dynamic scenes in order to consider the coherence assumption on surfaces that are truly in the dame space-time proximity.   \\  

\begin{figure}[h] 
\begin{center} 
\includegraphics[width=\columnwidth]{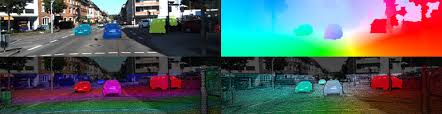}
\end{center}
\caption{Scene Flow Understanding Based on Video \cite{ObjectSceneFlowForAutonomousVehicles}.} 
\label{SceneFlowUnderstandingBasedOnVideo}
\end{figure}

Menze \textit{et al.} \cite{ObjectSceneFlowForAutonomousVehicles} reasoned about decomposing the scene into independently moving indexed objects and using a minimal representation of each one by its rigid motion parameters and each superpixel by a 3D plane. The use of the minimal representation leads to a discrete-continuous CRF and reduces the data into pairwise potentials between superpixels and objects. The 3D location and the 3D flow of each object that delimited by a superpixel in the scene is found by associating every superpixel with a 3D plane variable and a pointer to an object comprising its rigid motion. Then, given consecutive frames of the reference frame, the developed object scene flow infered the 3D geometry of each superpixel, the association to corresponding objects as well as the rigid body motion of each object, as shown in Fig. \ref{SceneFlowUnderstandingBasedOnVideo}.    

While in \cite{3DSceneFlowEstimationWithAPiecewiseRigidSceneModel}, the authors proposed to segment the input images by representing the dynamic scene as a collection of rigidly moving planes and then jointly recovering the geometry and the 3D motion when over-segmenting the scene. The developed piecewise rigid scene is intended to represent real world scenes with independent object motions rather than pixel-based representations like partially used in \cite{ObjectSceneFlowForAutonomousVehicles}. \\

As opposed to the two aforementioned research works that are considered as pure computer vision solutions, the authors in \cite{FeatureSpaceOptimizationForSemanticVideoSegmentation} attempt to create 3D reconstruction of dynamic scenes by achieving a long-range spatio-temporal regularization in semantic video segmentation, since both the camera and the scene are in motion. The developed idea is to integrate deep convolutional networks and CRF to perform sharp pixel-level recognition of boundaries of objects. To this end, the solution minimizes the distances between features associated with corresponding points in the scene and consequently optimizes the feature space that is used by the dense CRF.    

\subsection{Direct Perception Approaches for Autonomous Driving Systems}
The direct perception approach was initially introduced by \cite{DeepDrivingLearningAffordanceForDirectPerceptionInAutonomousDriving} to predict the driving decisions, instead of learning to map images containing different situations with the steering action or parsing and recognizing the scene and deciding the control actions of the vehicle. The developed model is built on top of a CNN that automatically derives indicators of situation of the road, such as the distance between the vehicle and the lane marking, angle of the vehicle relative to the road and distances to other surrounding vehicles. Additionally, the adopted and trained CNN learns to extract image features and makes meaningful predictions and description for scene understanding and autonomously drive the vehicle by the help of a simple controller.

\subsection{Behavior Reflex Approaches for Autonomous Driving Systems}
Behavior reflex approaches \cite{ALVINNAnAutonomousLandVehicleInANeuralNetwork} are drastically different from mediated perception approaches and direct perception approaches by constructing a direct mapping of the sensed input to the corresponding driving action that has been taken. One old technique is developed in \cite{NeuralNetworkPerceptionForMobileRobotGuidance} that generated a neural network to create a direct mapping from input images to corresponding steering angles in order to be directly applied to learn driving actions. To learn the model, the system uses the images and the actions representing the steering angles as the training data. It has been proven that this technique can handle lane navigation, but failed to realize accurate decisions in a complicated traffic scenario. \\

As opposed to the Autonomous Land Vehicle in a Neural Network (ALVINN) system developed in \cite{ALVINNAnAutonomousLandVehicleInANeuralNetwork} that uses fully-connected networks, the end-to-end DAVE-2 system proposed by NVIDIA in \cite{EndToEndLearningForSelfDrivingCars} uses a CNN that directly maps the raw pixels of images from cameras to the steering commands. DAVE-2 makes use of CNNs that offers more than pattern recognition, as it is trained to learn the whole processing pipeline, to find steering decisions based on training video data collected while driving from two cameras and coupled with left and right steering decisions.

\subsection{LIDAR Perception Approaches for Autonomous Driving Systems}

The work in \cite{LIDARBasedPerceptionSolutionForAutonomousVehicles} developed a solution based on LIDAR data only to explore the vehicle surroundings. This solution involves three consecutive phases; namely, segmentation, fragmentation detection and clustering, and tracking. The developed solution is built based on combining multiple artificial intelligence techniques and it is conceived for detecting and tracking objects of any shape that are going to be used by autonomous vehicles to find and understand the scene flow of its surroundings. 
\begin{figure}[h] 
\begin{center} 
\includegraphics[width=\columnwidth]{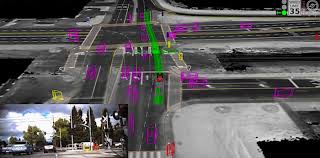}
\end{center}
\caption{LIDAR-Based Scene Understanding.} 
\end{figure}

\subsection{Mixed Mediated and LIDAR Perception Approaches for Autonomous Driving Systems}

\begin{figure}[h] 
\begin{center} 
\includegraphics[width=\columnwidth]{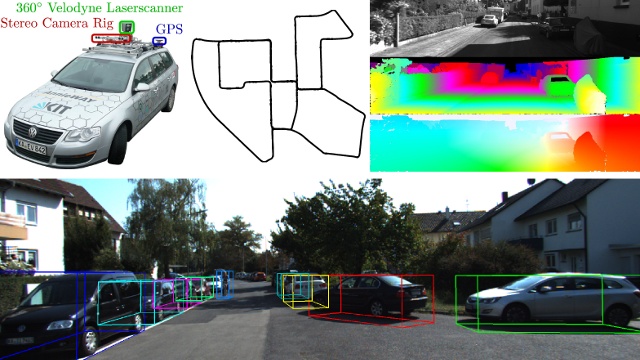}
\end{center}
\caption{KITTI Benchmark Suite for creating the Dataset \cite{KittiDatasetBenchmark}.} 
\end{figure} 

The autonomous driving platform developed in \cite{KittiDatasetBenchmark} presents an interesting and challenging real-world computer vision benchmarks that include stereo, optical flow, visual odometry, 3D object detection and 3D tracking of real time surroundings of autonomous vehicles.  The data includes the LIDAR and the camera recordings of driving sequences.
\begin{figure}[h] 
\begin{center} 
\includegraphics[width=\columnwidth]{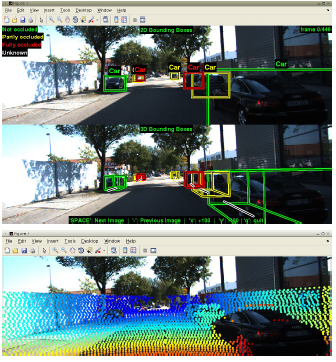}
\end{center}
\caption{Real-Time 3D Object Detection and Tracking.} 
\end{figure}

\section{Localization and Mapping For Self-Driving Vehicles}\label{SectionConclusions}
Traditional localization approaches such as GPS are inaccurate with a significant localization error in addition to the total unreliability because of the insufficient network coverage.   
Digital driving instructions have to be ultra-precise since driving requires steering commands to be in consensus with the deep-learning and mapping of the outdoor environment. \\
Consequently, fully AD systems opt to use  high-precision maps or also called HD maps to localize a vehicle in the space and within its surrounding objects in the scene. Automatically created high-precision decision maps put together various enriched multi-model sensed data and a pre-mapped environment that can be from raw LIDAR, Video or mixed sources.\\
As opposed to standard definition maps that are based on GPS positioning such as Google Maps and used for machine to human communications, HD maps are fundamentally different with a better resolution and are meant for machine-to-machine communications. Its relative localization systems are supposed to provide an approximate maximum localization precision of $25$ cm, in comparison with a $2$ to $10$ meter accuracy in GPS, without the need for external error detection and correction systems.\\
Major research attempts are discussed in this section to gain a better insight into the real practicality of SLAM \cite{SlamGeneralReference} for localization of self-driving vehicles through odometry \cite{OdometryGeneralReference} that uses data irrespective of sensing modality to approximate the changes in a vehicle's position over time.

\subsection{Visual Odometry for Vehicle Localization}
\begin{figure}[h] 
\captionsetup{singlelinecheck = false, justification=justified}
\begin{center} 
\includegraphics[width=\columnwidth]{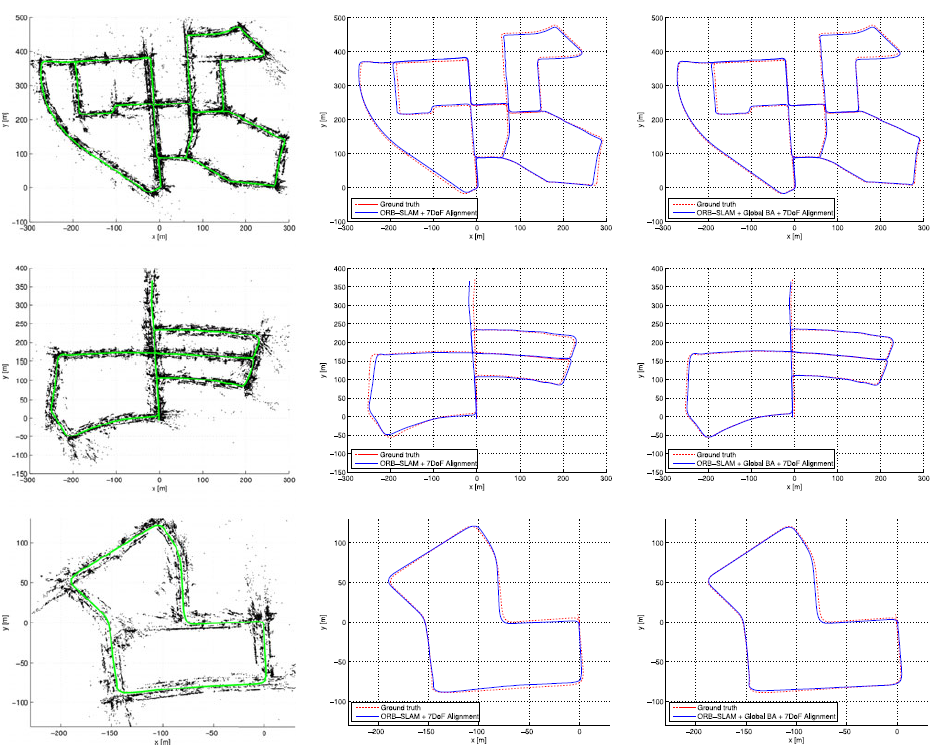} 
\caption{Sequences 00, 05, and 07 From the Odometry Benchmark of the KITTI dataset. (Left) Points and Keyframe Trajectory. (Center) Trajectory and Ground Truth. (Right) Trajectory After 20 Iterations of Full BA \cite{7219438}.} 
\label{orb-slam-final}
\end{center}
\end{figure}

Scanned roads by stereo cameras overlaid over high precision positional data produce a highly detailed representation and localization. 
In this context, Raul \textit{et al.} developed in \cite{7219438} a novel feature-based monocular SLAM system that requires creation of initial maps since depths may not be found from a single image frame in order to provide wide baseline loop closing detection and to serve motion clutter. The developed solution covers the tracking, mapping, relocalization and loop closing based on a novel strategy that selects specific key important points and keyframes to generate trackable maps that grow with scene content changes.\\
Every Map point holds the 3D position of the world coordinate system, the viewing direction consisting of rays joining the point with the optical center of the keyframes observing it, and a representative ORB descriptor holding the smallest hamming distance among all associated descriptors in keyframes where the same point is observed. \\
In addition, the proposed scheme ensures that every keyframe stores camera principal points and focal length, ORB features that were extracted from the frame independently of whether or not they are associated with map points and the camera pose that represents rigid body transformation of points from real-world coordinates to camera coordinates. Then the co-visibility graph, the essential graph, ORB descriptors extraction and the automatic map initialization are achieved dynamically in order to satisfy finding initial correspondences and making the parallel computation to find inliers between models.
Fig. \ref{orb-slam-final} shows points and keyframe trajectory, ground truth and trajectory and after many iterations of full bundle adjustment.

\begin{figure}[h] 
\captionsetup{singlelinecheck = false, justification=justified}
\begin{center} 
\includegraphics[width=\columnwidth, height=4cm]{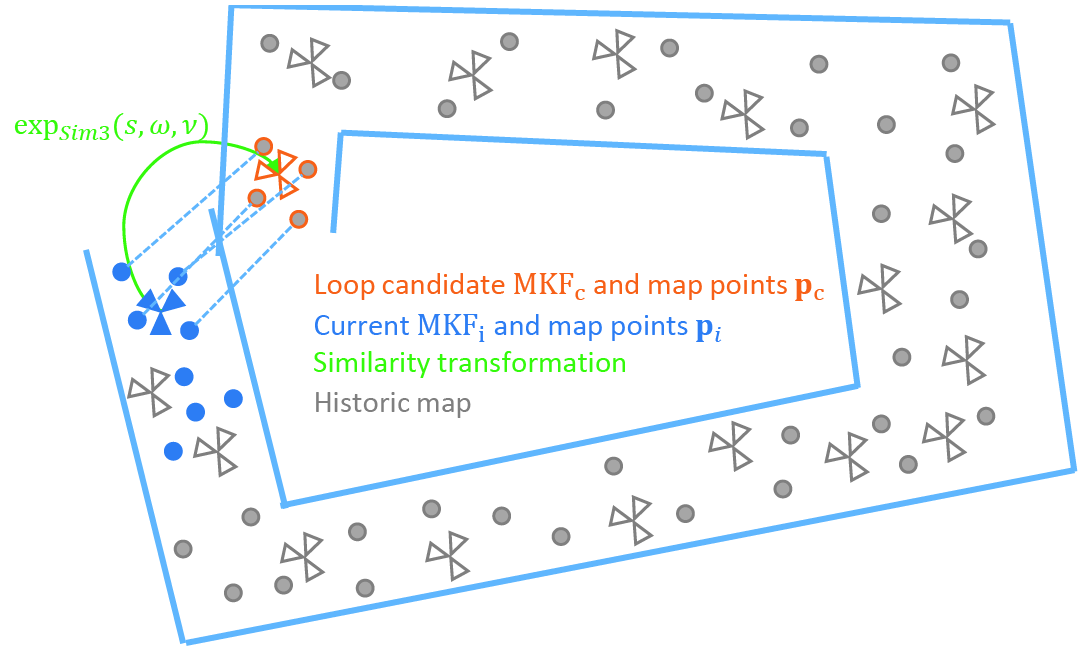} 
\end{center}
\caption{Loop Detection Example on an Uncoinciding Spatial Alignment of Local Map With Historic Map \cite{Urban2016MultiColSLAMA}.} 
\label{multiColPictureFinal}
\end{figure}

A continuous estimation of vehicle's current position and orientation depending on the observed environment from the camera is presented in \cite{Urban2016MultiColSLAMA}. It enables SLAM with rigidly coupled frames of Multi-Camera Systems (MCS) by taking advantage of the MultiCol Model and introducing: \\
\begin{itemize}
    \item The usage of Multi-Key Frames (MKFs) with identification and match points across them and mask outliers.  
    \item Hyper-graph formulation of MultiCol for different tracking and modeling pipeline than used in \cite{7219438} where graph edges connect two vertexes only and now edges can connect to any arbitrary number of vertexes.
    \item Advanced Multi-camera loop closing that searches eventual loop closures from newly presented MKF in order to figure out if a place has been visited. 
\end{itemize}
Depiction of the loop closing, applying similarity transformation to possible alignment of points and usage of the alignment error to correct the remaining MKF poses and points mapping by projection to map are generalized in Fig. \ref{multiColPictureFinal}. \\

\begin{figure}[H] 
\captionsetup{singlelinecheck = false, justification=justified}
\begin{center} 
\includegraphics[width=\columnwidth, height=5cm]{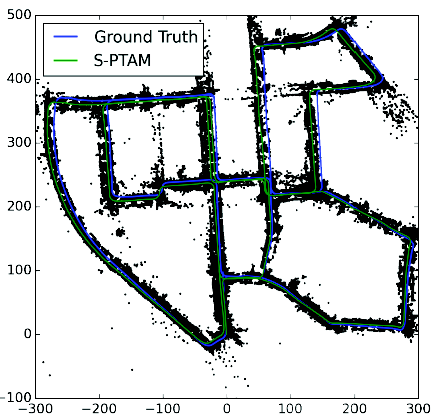} 
\end{center}
\caption{Estimated Trajectory Performed by S-PTAM on KITTI Dataset 00 Compared to the Ground Truth as well as the Generated Map \cite{7353546}.} 
\label{stereo_parallel_tracking}
\end{figure}

The authors in \cite{7353546} presented a stereo SLAM localization system called S-PTAM that uses only a stereo camera to avoid the monocular bootstrapping problem as in \cite{7219438}. To improve robustness, the authors used to enforce stereo constraints on the pose- and map- refinment algorithms and applied a maintenance process for each independent thread by running a map bundle adjustment refinement algorithm in local area to improve global consistency.\\
Fig. \ref{stereo_parallel_tracking} draws the estimated trajectory discovered by S-PTAM compared to ground truth and to generated maps while appiled to the sequence $00$ of the KITTI dataset. 

\begin{figure}[h]  
\captionsetup{singlelinecheck = false, justification=justified}
\begin{center} 
\includegraphics[width=\columnwidth]{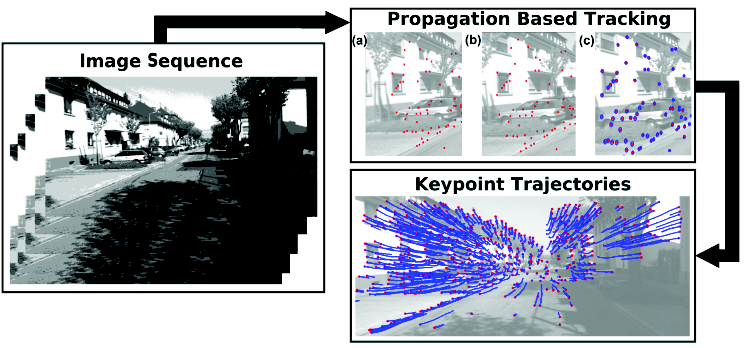} 
\caption{System Steps Modules From Sequence of Images to Propagation Based Tracking and Finally Keypoint Trajectories. The Keypoints (a) and (b) in Frame n-1 and Frame n, While Keypoints (c) is the Predicted Keypoints in Frame n+1 \cite{7535500}.} 
\label{keypoint_trajectory_Estimation}
\end{center}
\end{figure}

To achieve the creation of a reliable visual environment systems, the authors in \cite{7535500} present a scheme for tracking keypoint and estimation of egomotion and the structure of the environment from the trajectories of the selected keypoints.\\
Fig. \ref{keypoint_trajectory_Estimation} presents an overview of the pipeline of the solution that first consists of decomposing a sequence of images from a monocular camera setup into consecutive frames and applying a propagation based tracking method to find the 2D trajectories of the keypoints and not through finding keypoint correspondences like in \cite{7353546}.\\
The search of the keypoints matches from one frame is achieved by propagating the estimated 3D position of each keypoint from one frame to another and by making sure that the photometric consistency is verified. 

\subsection{LIDAR-Based Vehicle Localization}
In general, the rotational movement of LIDAR scanners and having the vehicle on the move too, make the registration and localization of the point cloud very difficult.


The authors in \cite{zhang2014loam} proposed a solution that uses odometry measurements to achieve low-drift and low-computational complexity by applying two algorithms. The first algorithm is applied to perform the odometry with low fidelity and high frequency for LIDAR velocity estimation. The second algorithm is applied to run with a lower frequency to perform both fine matching and registration of the point cloud data.\\
The LIDAR odometry algorithm uses last sweep point cloud, current sweep growing point cloud and the pose transform from last recursion as inputs parameters. It extracts feature points, finds its correspondence and applies motion estimation to a robust fitting. The mapping algorithm is used to match and register the reprojected point cloud to a specific time stamp, based on the undistorted point cloud and the pose transform that contains the LIDAR motion during the sweep generated by the LIDAR odometry algorithm. LOAM solution is ranked \#3, with a 0.70 \% , on the KITTI odometry benchmark for the average computed translational and rotational errors.

\begin{figure}[h] 
\captionsetup{singlelinecheck = false, justification=justified}
\begin{center} 
\includegraphics[width=\columnwidth]{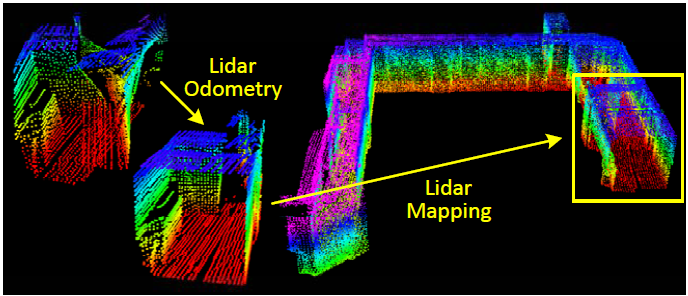} \label{lidar_odometry_and_mapping_final}
\caption{The Developed Solution Decomposes The Problem Into Two Parallel Algorithms: The Odometry Algorithm Used to Estimate the Velocity of LIDAR and to Correct Distortion in the Point Cloud, then, a Mapping Algorithm Matches and Registers the Point Cloud to Create a Map \cite{zhang2014loam}.} 
\end{center}
\end{figure} 


The authors in \cite{chong2013synthetic} developed a novel idea of synthetic 2D LIDAR in order to achieve precise vehicle localization on a virtual 2D plane. The presented Monte Carlo Localization to estiamte the position of the vehicle replies on the synthetic LIDAR measurements and odometry information. A demonstration of the accuracy and robustness of this approach is demonstrated through carrying out real-time localization of the driving test in the NUS campus area. 
\begin{figure}[h] 
\captionsetup{singlelinecheck = false, justification=justified}
\begin{center} 
\includegraphics[width=\columnwidth, height=5cm]{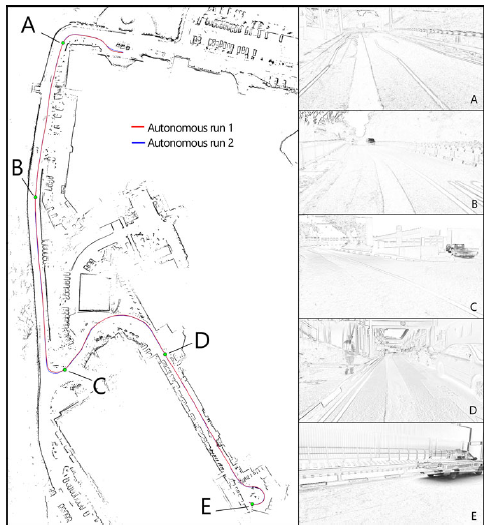} \label{synthetic_2d_lidar_final}
\end{center}
\caption{Autonomous Navigation With Synthetic Virtual LIDAR. Images on the Right from Top to Bottom Correspond to Visual Validation of Localization Repeatability From Checkpoint A to E \cite{chong2013synthetic}.}  
\end{figure}

In \cite{brenner2010vehicle}, the authors studied an accurate and reliable vehicle localization system by using the landmarks acquired from LIDAR mapping systems. Since the standard Global Navigation Satellite System (GNSS) solutions lacks high reliability, the developed solution relies on relative localization using LIDAR sensors and a pre-mapped environment. Consequently, the creation of detailed environment maps is a necessary step to produce  accurate localization. The obtained landmark pairs are then associated with each others using an estimation approach, which leads to accurate position estimations of the robotic vehicle.

\begin{figure}[h] 
\captionsetup{singlelinecheck = false, justification=justified}
\begin{center} 
\includegraphics[width=\columnwidth, height=4cm]{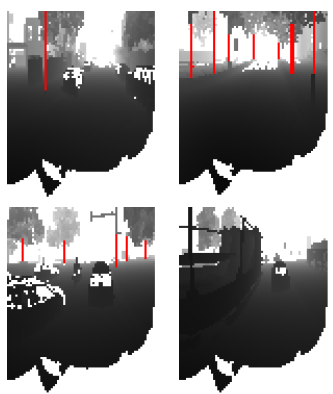} \label{pole_detection_final}
\end{center}
\caption{ Pole Detection Results With Black is Near, White is Far (or error), and Detected Poles are Marked in Red \cite{brenner2010vehicle}.} 
\end{figure}

Additionally, more advanced methods were incorporated, such as using dead reckoning based on multiple matching successive scans as presented in SLAM approaches. Alternatively, the authors discussed other landmark incorporating examples like planar patches. 
Then, they investigated in details the matching algorithm with the RMS variations, and the weak configuration points that were automatically detected. This yield smaller error bounds that met the requirement of the  design in terms of reliable and automatic association procedure.

\begin{figure}[H] 
\captionsetup{singlelinecheck = false, justification=justified}
\begin{center} 
\includegraphics[width=\columnwidth, height=4cm]{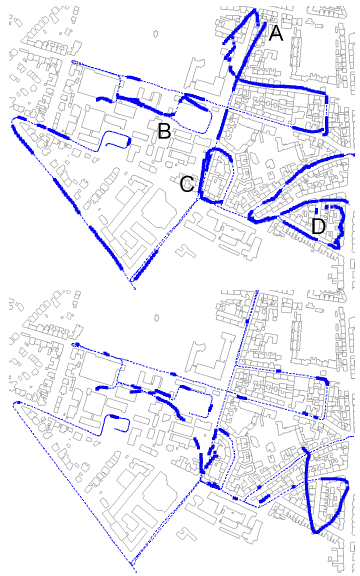} \label{comparison_gps_lidar_odometry_final}
\end{center}
\caption{GPS Trajectory on Top and Result of Matching on Buttom \cite{brenner2010vehicle}.} 
\end{figure}

\subsection{Combined Visual-LIDAR Odometry for Vehicle Localization}

It is widely used to utilize the dense point clouds with measures of surface reflectivity obtained by LIDAR scans to obtain localization results of centimeter-level accuracy. However, this comes at a prohibitive cost in terms of sensor suites and complex state-of-the-art localization techniques. This is why, the authors in \cite{wolcott2014visual} investigated a cheaper model that finds comparable localization accuracy using cameras.\\ The developed model is built to localize a single monocular camera within an already created 3D ground map from a surveying vehicle equipped with 3D LIDAR scanners. To realize that, a GPU is used to create several synthetic views of the visited environment in order to achieve maximal normalized mutual information between the GPU and the real camera measures.
\begin{figure}[H] 
\captionsetup{singlelinecheck = false, justification=justified}
\begin{center} 
\includegraphics[width=\columnwidth]{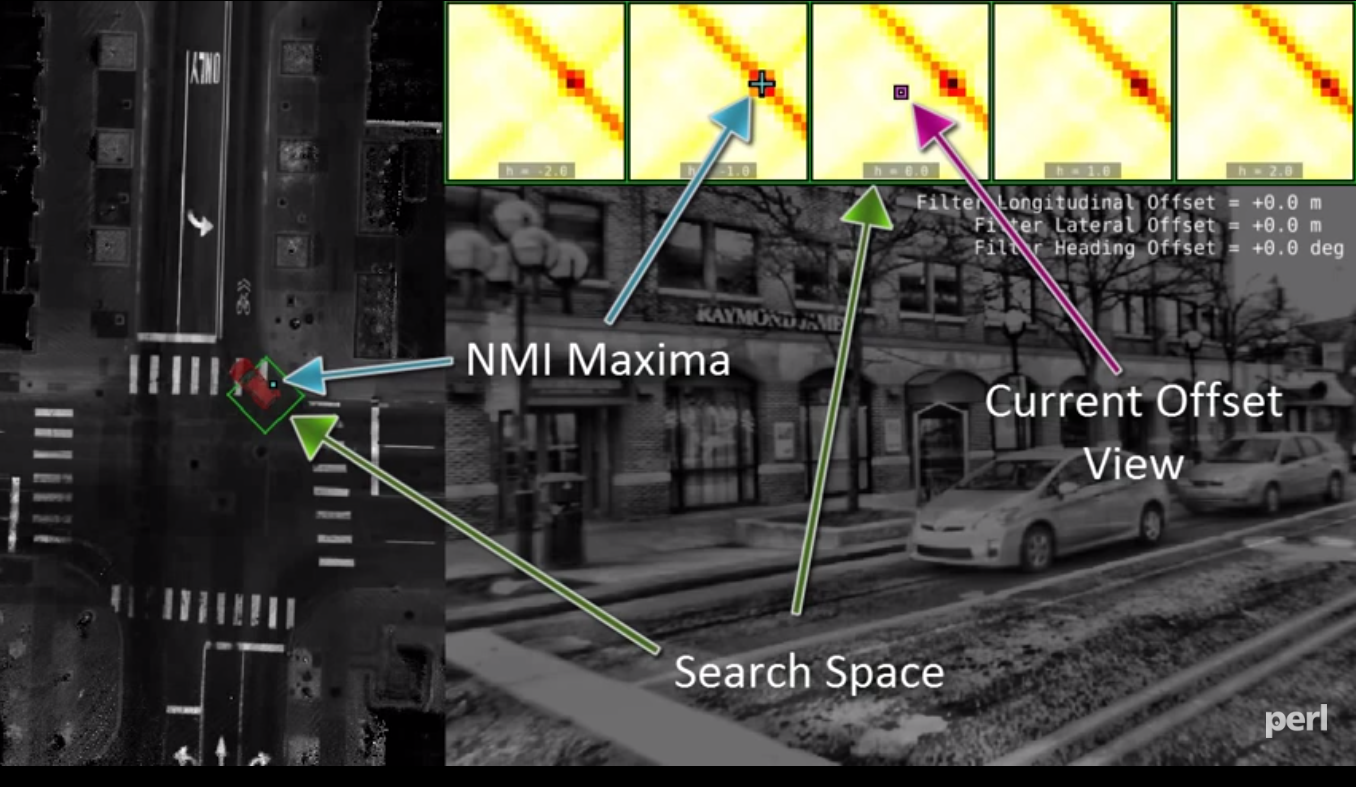} \label{Visual_Localization_Final}
\end{center}
\caption{ 
Localization Within LIDAR Maps with Bird’s Eye View on Left, Normalized Mutual Information Cost Surface On Top and Alpha Blended Camera$/$synthetic View on Bottom \cite{wolcott2014visual}.} 
\end{figure}


In \cite{zhang2015visual}, the authors present a general framework for combining visual and LIDAR odometries in a fundamental and novel method. The method shows improvements in performance over the state of the art, particularly in robustness to aggressive motion and temporary lack of visual features. The proposed on-line method starts with a visual odometry to estimate the ego-motion and to register point clouds from a scanning LIDAR at a high frequency but low fidelity. Then, scan matching based LIDAR odometry refines the motion estimation and point cloud registration, simultaneously. In addition to the comparison of the motion estimation accuracy, the authors evaluated the robustness of the method when the sensor suite moves at a high speed and is subject to significant ambient lighting changes. \\ 
The overall system is divided into two sections. The visual odometry section estimates frame to frame motion of the sensor at the image frame rate, using visual images with assistance from LIDAR clouds. In this section, the feature tracking block extracts and matches visual features between consecutive images. \\
The depth map registration block registers LIDAR clouds on a local depthmap, and associates depth to the visual features. The frame to frame motion estimation block takes the visual features to compute motion estimates.
When applied to the KITTI odometry benchmark for assessing the translational and rotational errors, V-LOAM achieves 0.75\% of relative position drift and is ranked \#2. 
\begin{figure}[H] 
\captionsetup{singlelinecheck = false, justification=justified}
\begin{center} 
\includegraphics[width=\columnwidth]{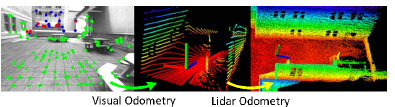} \label{visual_lidar_odeometry}
\end{center}
\caption{Motion Estimation and Mapping are Achieved Using a Monocular Camera Combined with 3D LIDAR \cite{zhang2015visual}.} 
\end{figure}

Visual odometry can be augmented with depth information as provided by RGB-D cameras, or from LIDARs associated with cameras. However, such depth information can be limited by the sensors, leaving large areas in the visual images where depth is unavailable.\\
The authors \cite{zhang2014real} propose a method to utilize the depth, even if sparsely available, in recovery of camera motion. In addition, the method utilizes depth by triangulation from the previously estimated motion, and salient visual features for which depth is unavailable. The core of the proposed method is a bundle adjustment that refines the motion estimates in parallel by processing a sequence of images, in a batch optimization. \\
The method in \cite{zhang2014real} for the three sensor setups, one using an RGB-D camera, and two using combinations of a camera and a 3D LIDAR. The presented method is rated \#2 on the KITTI odometry benchmark irrespective of the sensing modality, and is rated \#1 among visual odometry methods. \\

\begin{figure}[H] 
\captionsetup{singlelinecheck = false, justification=justified}
\begin{center} 
\includegraphics[width=\columnwidth]{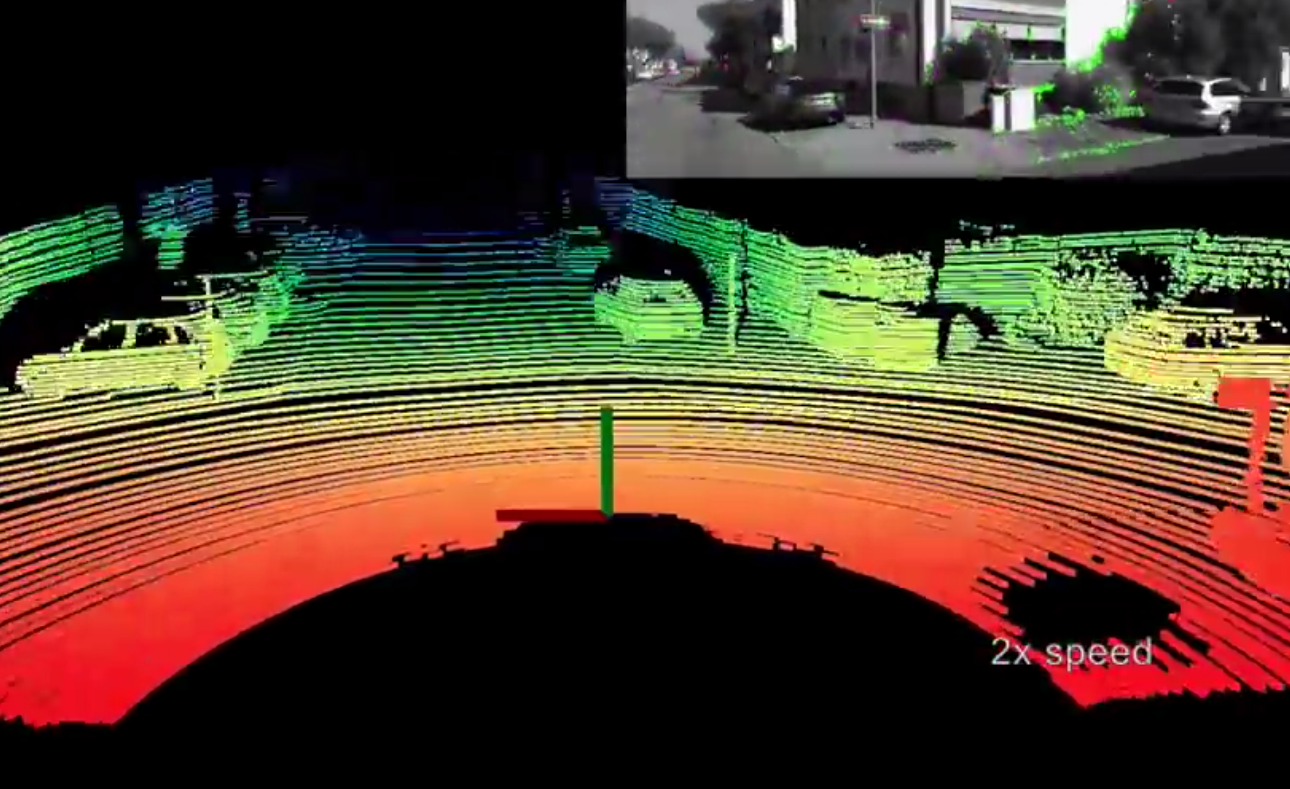} \label{real_time_depth_estimation_monocular_final}
\end{center}
\caption{Point Cloud Perceived by Velodyne Scanner on KITTI Dataset \cite{zhang2014real}.} 
\end{figure}

In \cite{levinson2007map} the authors developed a complete solution for high-accuracy localization of vehicles on the move using urban environments' maps. The presented solution aims to generate high-resolution environment maps by integrating GPS, IMU, wheel odometry, and LIDAR point cloud data obtained from the vehicle's recorded data. The solution allows the localization of the moving vehicle within the maps by mainly using a particle filter method for calculating the correlation of LIDAR measurements with the map. \\
The map is created by the reconstruction the 3D structure of the environment using infrared reflected laser beams of the LIDAR. More precisely, the experimental results relative accuracy are better by more than an order of magnitude than the ones using traditional methods (e.g., GPS, IMU and odometry). Concretely, the presented solution achieved a reliable real-time localization within $10$ centimeter range accuracy during bad whether and without GPS services, as opposed to the non accurate outdoor localization research work based only on GPS. 
\begin{figure}[H] 
\captionsetup{singlelinecheck = false, justification=justified}
\begin{center} 
\includegraphics[width=\columnwidth, height=4cm]{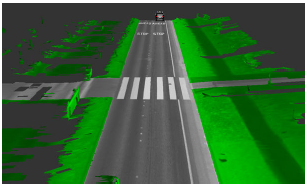} \label{map-based-precision-final}
\end{center}
\caption{ Visualization of the Ground Plane Extraction with the Measurements that
Coincide with the Ground Plane Are Retained and in Green All Others Are Discarded \cite{levinson2007map}.} 
\end{figure} 

\textbf{{\normalsize PART 4: TOWARDS VCs MEETING AUTONOMOUS VEHICLES}} 

In full alignment with the USDoT’s ITS-JPO vision stated in the RFP, we innovatively blend connected vehicle technology with autonomous vehicle systems. These systems will leverage the power of V2V and V2I communications, to significantly enrich the vehicle understanding of its surroundings and collectively perform driving decisions given the exchanged dynamics, movements, and intents of all vehicles. \\

In addition to the communications impact, VCs are expected to play a major role in supporting the limited capabilities of vehicles with on the move add-on functionalities. The VC will contribute to creating solid learning formulations to build more accurate 3D scenes of the vehicle surroundings, using the acquired information from multiple input modes, such as cameras, LIDARs, and exchanged V2V/V2I messages. \\

The targeted multimodal solutions will insert all the information acquired by the input modes into one mapping and learning setting, which is very likely to be the revolutionary HD live maps. It encodes one accurate 3D space representing the surrounding scene including V2V-enabled out-of-sight vehicles and V2I-conveyed sign/signal information, 2D images with rich texture description and accurate object depth in 3D LIDAR point clouds. 
\section{Post Alignment and Fusion}
Maalej \textit{et al.} presented in \cite{VanetsMeetAutonomousVehicles} a multimodal framework for object detection, recognition and mapping based on the fusion of stereo camera frames, point cloud Velodyne LIDAR scans, and V2V exchanged BSMs over DSRC. Based on the adapted Darknet's CNN, the pixel-wise adjacency coordinates of moments are derived from the bounding boxes of recognized objects of KITTI frames. In addition, objects are retrieved from point cloud scans to the same camera capture. The generated BSMs corresponding to the vehicles from LIDAR have position (x,y,z) triplet as a replacement to the real-world positioning from (Latitude, Longitude, Elevation). The authors presented a semi-supervised manifold alignment technique to achieve camera-LIDAR and camera-V2V mapping of recognized objects that have the same underlying manifold.   \\

The post-alignement consists of matching correpsonding pairs of recognized objects to reconstruct a 3D world as a virtual map surrounding the self-driving vehicles. Finally, the presented enriched points are the ones that were not mapped as they might correspond to out-of-sight objects, or BSMs of distant cars, etc. \\

Alternatively in \cite{CloudToCarMappingSystem}, the idea of a cloud-to-car mapping system is presented by Archer Software to have from the cloud the most updated details about road networks like lane curvature, roadway profile, static positions of traffic signs and then recognized objects from the car itself are post-added in order to achieve automatic maneuvering.  \\

In \cite{kumar2012cloud}, the authors present a novel cloud-assisted design for autonomous driving that is called Carcel, where all the sensed data from self-driving cars and from static roadside sensors are post-aggregated and managed in the cloud. Precisely, the cloud is responsible for conveying obstacles to vehicles after analyzing the received information, which means that the cloud will assist vehicles on roads to make decisions to avoid obstacles that are not necessarily sensed by their sensors. Moreover, the longer range information that the cloud builds about the traffic patterns enables more efficient path planning. This method is not realistic due to the huge amount of data that is sent from and to the cloud, which delays the steering decisions arriving to vehicles. 

\section{Pre-fusion between VC and AD Data Modalities}

As opposed to the post-fusion and mapping systems, the pre-fusion systems do not rely on having seperate recognized data from sensors or cloud that are mapped together. For example, Mobileye's REM system \cite{bmwAndMobileyeToCrowdsource} relies first on a pre-existing rich HD maps that are supplemented with to-the-minute data from real-world driving. An advanced faster version for accurate real-time mapping of data is currently under development.   

An issue with the pre-fusion of cloud data of in-vehicle sensors is the threat of cyber attacks which may lead to a disastrous reconstruction of surrounding world and consequently to a failing AD system. As detailed in \cite{GoogleKeepsSelfDrivingCarsOffline} having self-driving vehicles relying on offline system might be the solution to hinder hackers.  \\

More simplistic and secure scheme that does not rely on incoming cloud data is developed in \cite{MultiView3DObjectDetectionNetwork}, where a Multi-View 3D networks (MV3D) is developed to guarantee accurate 3D object detection applied to autonomous driving scenarios. The developed sensory-fusion framework is used to predict the orientation of 3D bounding boxes representing an object by taking both raw RGB images and LIDAR point cloud without applying any prior recognition. The generated candidates of 3D boxes for localization and detection in the KITTI benchmark outperforms the up-to-date systems by approximately 25\% and 30\%. 

Similarly in \cite{3DObjectDetection}, the 3D point cloud and 2D front view images are fused via a CNN with the usage of a Region Proposal Network (RPN) that is used in the network's multiple layers to generate proposal region objects in the front view.
\section{Cooperative and Coordinated Autonomous Driving}
In order to fully leverage the autonomy of self-driving vehicles, it is obvious that developing intelligent systems built on top of coordinated autonomy would guarantee additional safety with more trusted manoeuvres, reduce fuel consumption and balanced network traffic while guaranteeing a cooperative dispatching of vehicles depending on time and positions. In this context many ongoing research projects on vehicles coordination are being developed.\\

EPFL's research team working on the AutoNet2030 \cite{epflCooperativeDriving} presented in the design of distributed algorithms among vehicles with different levels of intelligence on borad ranging from fully autonomous to standard vehicles for cooperation among vehicles. The demonstration included an autonomous truck, an autonomous vehicle and an ordinary car equipped with a human machine interface that was able to satisfy cooperatively a merging scenario on a highway by offering enough space to a vehicles intending to merge between the two others. The robustness of the algorithms allows its deployability even before having all the vehicles being autonomous. \\

Another key feature of coordinated driving is being implemented by the authors in \cite{decentralizedCooperativeDriving} consisting of a decentralized trajectory coordination for autonomous vehicles without the usage of widely applied synchronization points that force all vehicles to pass by synchronously. The developed system assumes that a unique priority is assigned to every vehicle and each one periodically informs other vehicles about its planned future trajectory through V2V communications. Consequently, every vehicle will be responsible for maintaining the possible trajectory that will be collision-free with regard to other vehicles with higher priority. The collision-free path is obtained by the trajectory planned that takes into account space, time and a pre-designed network of paths while modeling higher-priority vehicles when moving. A real time scenario validating the developed  method is presented where every conflict between desired trajectories triggers automatic conflict resolution.
\section{Examples of Integration Systems}

\begin{figure}[H] 
\captionsetup{singlelinecheck = false, justification=justified}
\begin{center} 
\includegraphics[width=\columnwidth]{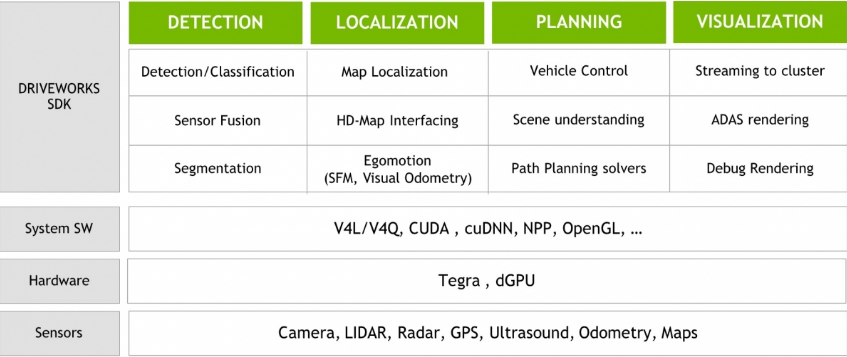} 
\end{center}
\caption{DriveWorks SDK Architecture \cite{nvidiaPX}.} 
\label{NVIDIADRIVEWORKS}
\end{figure}

NVIDIA's Drive PX \cite{nvidiaPX} AI car computer for self-driving vehicles is considered as a pioneering effort among all the computing platforms for real-time scene understanding, HD mapping, energy efficiency, automated highway driving, accelerated deep learning, sensor fusion, HD mapping, etc. Its GPU data parallel architecture accelerates the perception task of detecting and classifying objects using deep neural networks that are pre-trained in cloud datacenters and deployed in the vehicle. \\ 
In addition NVIDIA's recently developed Co-Pilot \cite{nvidiaCo-Pilot} is considered as an extra AI system built toward fully autonomous systems that is either driving the vehicle or looking out for the driver. In other words, for tricky traffic, changing roads without up-to date maps or highly fuzzy uncertain perception then the Co-Pilot makes sure to increase passenger driver awareness with surrounding environmental awareness. It notifies the driver about objects that she might need to be concerned about, increase the cautiousness of driver about where to look by focusing on the driver's face recognition, head tracking, gaze tracking, lips reading, etc. \\
NVIDIA's DriveWorks \cite{NvidiaDriveWorks} is a powerful Software Development Kit (SDK) for researchers and industry to implement any level of autonomous driving accelerated on Drive PX. Its layered architecture and run-time pipeline framework is detailed in Fig. \ref{NVIDIADRIVEWORKS} and offers the following: 
\begin{itemize}
    \item Detection Libraries used for various types of sensors processing, fusion for objects segmentation, detection and classification.
    \item Localization Libraries used for odometry mapping, map localization and HD maps interfacing.
    \item Planning Libraries to ensure functions of vehicle automatic control, overall scene understanding and intelligent path planning.
    \item Visualization Libraries offering applications for cluster display streaming, ADAS and debug rendering.
\end{itemize}

\begin{figure}[H] 
\captionsetup{singlelinecheck = false, justification=justified}
\begin{center} 
\includegraphics[width=\columnwidth]{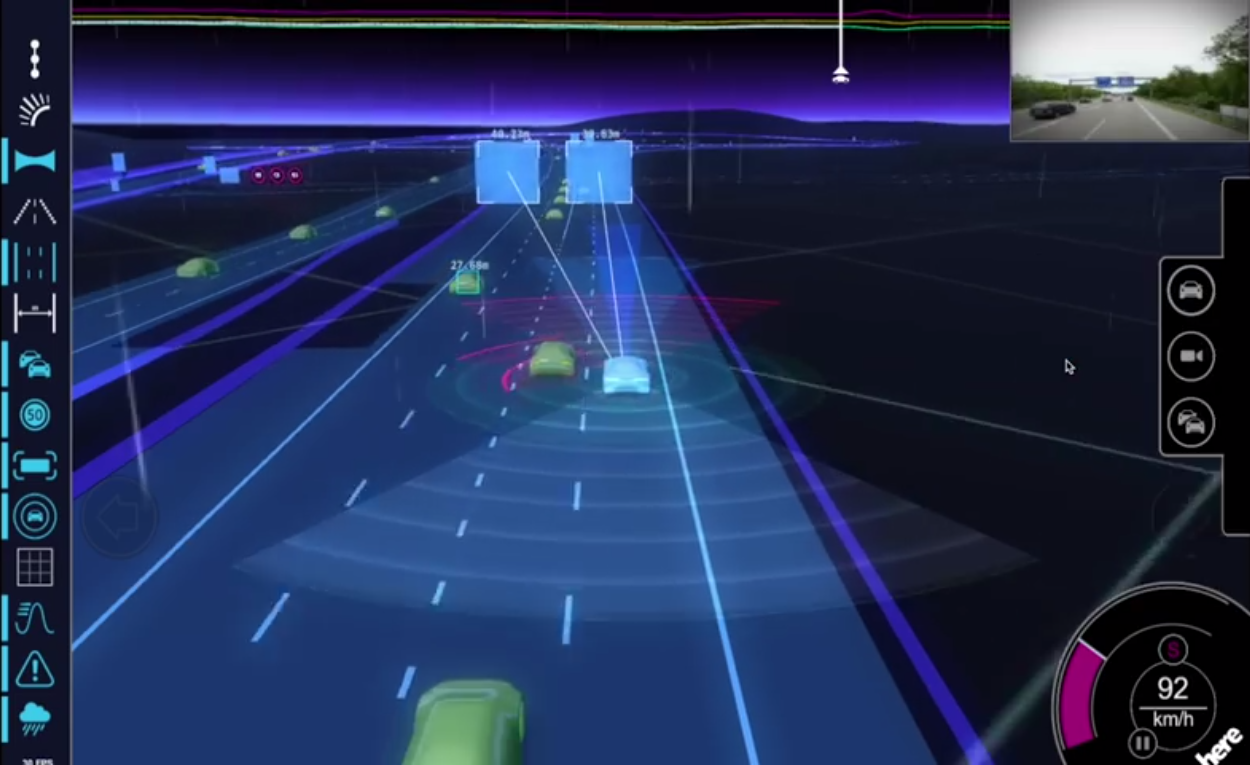} 
\end{center}
\caption{Visualization of HERE's Live HD Map.} 
\label{hereHdMaps}
\end{figure}

Partnering with NVIDIA on AI technology, HERE \cite{HereCitation} is moving forward to accelerate its HD Live Maps \cite{HereHdMaps} by using DRIVE PX 2 for its self-driving vehicle solutions that are delivered as a cloud service for a digital mapping content. The HD Live Maps is a layered complex solution, visualized in Fig. \ref{hereHdMaps}, can be depicted as:
\begin{itemize}
    \item Lane Model: is considered as the key model of the HD maps since it provides a highly accurate representation of the road networks with different lanes' separation and roads' delimiters, all mapper with a 10 to 20 cm precision.
    \item Localization Model: allows finding the exact position of the vehicle in the lane where it is driving on the HD Map. Interestingly, it allows cars to be map makers since as long as they are being on the road, they collect data about the perceived environment and provide it to the cloud which aggregates all received information to update the maps.  
    \item Activity Layer: plays a major role in understanding dynamic events while driving on the road network such as heavy traffic, accidents, road closures, and other dynamic events that may impact the driving strategies and paths of the vehicle.  
    \item Analytic Layer is used to adapt to tricky scenarios on how to drive by looking at various driving styles that inform the vehicles on how to behave under similar conditions.
\end{itemize}
As an illustration, HERE HD Live Maps offer intelligent complete solutions needed to pave the road to raise the trust in the commercial sustainability and reusability of automated driving systems by vehicle manufacturers for automated driving solutions.    

With cameras, RADAR, LIDAR, unreliable GPS and more sensors, Waymo \cite{Waymo} as formerly known as the Google self-driving car has achieved self-driven 3 million miles by using complex heavy computation mapping and fusion of sensed data while using minimal access to cloud-based services. Up to now, the self-driving is not commercialized yet even though there have been different prototypes varying from modified traditional vehicles requiring human inputs to vehicles without steering wheels.     \\


Tesla's Autopilot \cite{TeslaAutopilot} is considered as the most popular full standalone hardware with software solution for autonomous cars that is being actively updated and refined as long as the number of driven miles get higher, more than 1.3 billion miles of data \cite{Tesla1.3BiliionsMiles}. All cars built by Tesla after 2014 are equipped with hardware autopilot that are capable of receiving updates from the cloud of the core software for self-driving, required drivers for the sensors, etc. \\
With its particular technology that relies on 8 cameras, surrounding both Model S and X, along with twelve ultrasonic sensors, RADAR and GPS, the Autopilot creates a 3D view of the vehicle's surrounding in every direction that is beyond human access. The enhanced Autopilot offers new capabilities varying from simple speed to traffic condition adjustment to complex lane changing without driver's input. \\

Despite these new technological advancements and market solutions for autonomous driving, most of the aforementioned solutions have at certain places, moments or situations low confidence of driving autonomously. With such low confidence, self-driving systems require the driver to take-over the commands for driving. \\

\textbf{ {\normalsize PART 5: OPEN CHALLENGES AND CONCLUSION}} \\ 
In this part, we analyze the challenges and future trends that need to be solved before having a wide popularity of infotainment applications over VC and safety services in VANETs. Equally important, the full transition from advanced driver assisting to autonomous systems requires many technologies ranging from HD maps to fail-proof software to be more mature and reliable before such vehicles hit the roads. \\

\section{Open Challenges}

The lack of a widely used and standardized wireless communications technology in VANETs is one the most challenging problems blocking the expansion of VCs. Existing state-of-the-art research on communications would satisfy the safety aspect required for V2V safety systems, but does not meet the need of the autonomous driving connectivity. \\ 

5G might be the complete communications technology that will ease transforming VCs and automation systems from a vision into a widely reasonably affordable reality. Consequently, the challenges facing 5G connections float on the surface as a disruptive factor for VCs exploitation. Similarly, knowing that self-driving vehicles are making Big Data even bigger, high speed connectivity like 5G would be the solution to tap into faster networks and to offload to the cloud the several hundreds of mega bits created per second. \\

The security issues in both the communications and the automation systems would be definitely areas of study of the standardized complete systems when developed. Additionally, sooner or later VCs are supposed to manage the microdatacenters mounted in vehicles, communications base stations and the traditional VC architectures in the hope that they will revolutionize various systems to be taken advantage of in modern vehicles.    \\

The existing low level vehicles' automation has not led to fully self-driving complete systems yet. This is due to the many open unsolved problems or existing solutions that need more refinements and improvements. \\
To put it another way, it is obvious that owning the most intelligent accurate HD maps will certainly own the future of self-driving vehicles, independently of the perception, planning and command technologies. That is the virtual-world map of streets, which allow the vehicles to accurately map themselves as if they are on well-known empty streets and add detected dynamic objects later. \\ 

Moreover, new sophisticated and cheaper sensors should be developed to make the business model of self-driving vhicles more attractive. For example, the spinning Velodyne LIDAR offers great 360 degrees visibility and accurate depth information, but comes at a high price tag and huge amount of created data that makes processing so complex.
Recently announced solid state LIDARs by Quanergy \cite{quanergySolidStateLidar} and Velodyne \cite{VelodyneSolidStateLidar} are expected to hit the market with few hundred US dollars in comparison to several thousand US dollars for the spinning LIDARs. \\

Some problems facing Spinning-LIDARs along with a comparison with solid state LIDAR are presented in \cite{SelfDrivingCarsSpinningLaserProblem} including durability, precision, stability, etc.
Fail-safe real-time software used as components in the autonomous driving systems are far from being completely developed and validated. These pieces of software may include failing functionalities for:
\begin{itemize}
    \item objects detection, classification and tracking.
    \item sensors fusion and vehicle mapping.
    \item deep scene understanding and path planning.
    \item control and steering systems.
\end{itemize}


\section{Conclusion}\label{SectionConclusions}
Self-driving vehicles are known as the mother of artificial intelligence systems and the one with most promising business models. The emerging idea of VC is very promising too, aiming to improve the driving experience by using various types of infotainment applications, and to increase the vehicles safety by using beaconing applications and supporting extra additional over the air data for self-driving systems. \\
Moreover, many challenges and issues related to the vehicular cloud and vehicular network management especially with the fast changing topology have been presented. Key challenges to enable VC and AD systems to be coherent and reinforce the desired functionalites either related to the driving commands or the driving experience are also presented. We studied the various complex components that need to be built, integrated and fit together to guarantee an autonomous driving vehicle. The sensing, perception, decision and actuation layers come with many challenges to be deployed, tested and then put into real-world systems. The challenges that involved in these layers are presented in this paper.
We also presented an overview of the existing communication technologies, VANET MAC protocols, algorithms for VANET network coalition, reliability of safety and non-safety applications, Vehicular cloud selection based on cost and QoS, all used to bridge reliable communications, VC and AD systems.


\section*{List OF Acronyms}

\begin{abbreviations}
\item[IOV] Internet of Vehicles
\item[VC] Vehicular Cloud
\item[VCC] Vehicular Cloud Computing
\item[AD] Autonomous Driving
\item[AI] Artificial Intelligence
\item[V2V] Vehicle-to-Vehicle
\item[V2I] Vehicle-to-Infrastructure
\item[VANETs] vehicular ad-hoc networks 
\item[NSF] National Science Foundation
\item[USDoT] US Department of Transportation
\item[OBU] Onboard unit 
\item[DSRC] Dedicated Short Range Communication   
\item[LTE] Long Term Evolution 
\item[WAVE] Wireless Access for Vehicular Environment  
\item[CCH] Control Channel   
\item[SCH] Service Channel   
\item[MAC] Media Access Control   
\item[SLAM] Simultaneous localization and mapping  
\item[BSM] Basic Safety Message 
\item[NPRM] Notice of Proposed Rule-Making 
\item[NHTSA] National Highway Traffic Safety Administration 
\item[SI]  Synchronization Interval  
\item[QoS] Quality of Service  
\item[FCC]  Federal Communication Commission   
\item[AC] Access Categorie  
\item[WSM]  WAVE Short Message  
\item[PHY]  Physical Layer 
\item[CCHI] Control Channel Interval   
\item[SCHI] Service Channel Interval   
\item[V2X]  Vehicle to Any 
\item[CAM] Cooperative Awareness Messages  
\item[MBMS] Multimedia Broadcast and Multicast Services
\item[eMBMS]  evolved MBMS   
\item[MANETs] mobile ad hoc networks   
\item[TDMA] Time Division Multiple Access   
\item[FDMA] Frequency Division Multiple Access  
\item[CDMA] Code Division Multiple Access   
\item[POA] Point of Attachment 
\item[VHO] Vertical Handoff   
\item[AODV]  Ad-hoc On Demand Distance Vector 
\item[DSDV] Destination-Sequenced Distance-Vector  
\item[DSR] Dynamic Source Routing  
\item[GPSR] Greedy Perimeter Stateless Routing  
\item[GPCR]   Greedy Perimeter Coordinator Routing
\item[GeoDTN] Geographic Delay Tolerant Network Navigation 
\item[+Nav] with navigation
\item[STAR] Street and Traffic Aware Routing  
\item[A-STAR] Anchor-based STAR  
\item[IEGRP] Infrastructure Enhanced Geographic Routing Protocol
\item[T-TSG] Time Stable Geocast   
\item[PREQ]  Propagation of Request  
\item[GeoSPIN] Geocast Spatial Information   
\item[IVG] Inter-Vehicles Geocast
\item[RTS]  Request To Send 
\item[CTS]  Clear To Send   
\item[ACK]  Acknowledgement
\item[SSVW] Stop Sign Violation Warning
\item[RCVW] Railroad Crossing Violation Warning
\item[SWIW] Spot Weather Information Warning 
\item[OVW] Oversize Vehicle Warning
\item[RSZW] Reduced Speed Zone Warning
\item[FCW] Forward Collision Warning  
\item[LTA]  Left Turn Assist
\item[IMA] Intersection Movement Assist
\item[BSW] Blind Spot Warning
\item[LCW] Lane Change Warning
\item[RSS] Received Signal Strength
\item[BEC] Back-End Cloud
\item[INS] Inertial Navigation Sensors
\item[VM] Virtual Machine
\item[MDP] Markov Decision Process
\item[CC] Cloud Controller
\item[WRI] Wireless Roadside Inspection
\item[ISP] Internet Service Provider
\item[CSP] Cloud Service Provider
\item[VCSP] Vehicular Cloud Service Provider
\item[MCC] Mobile CLoud Computing 
\item[CDAR] Cloud Directory Application Repository
\item[DOS] Denial of Service
\item[NaaS] Network as a Service
\item[STaaS] Storage as a Service
\item[CaaS] Cooperation as a Service
\item[INaaS] Information as a Service
\item[SaaS] Software as a Service
\item[PaaS] Platform as a Service
\item[FSL] Full Scene Labeling
\item[MRF] Markov Random Fields
\item[CRF] Conditional Random Field
\item[CNN] Convolutional Neural Network
\item[ALVINN] Autonomous Land Vehicle in a Neural Network 
\item[GPS] Global Positioning System
\item[MCS] multi-camera systems
\item[MKFs] Multi-Keyframes
\item[GNSS] Global Navigation Satellite System 
\item[MV3D] MultiView 3D networks 
\item[RPN] Region Proposal Network 
\item[SDK] Software Development Kit
\end{abbreviations}

\bibliographystyle{ieeetr}
\bibliography{BibliographyFinal}

\end{document}